\documentclass[12pt,english]{article}
\usepackage[T1]{fontenc}
\usepackage[latin1]{inputenc}
\usepackage{a4wide}
\usepackage{subfigure}
\usepackage{graphicx}
\usepackage{amssymb}

\makeatletter

\newcommand{\noun}[1]{\textsc{#1}}
\providecommand{\tabularnewline}{\\}

\makeatletter
\renewcommand{\theequation}{\hbox{\normalsize\arabic{section}.\arabic{equation}}}
\@addtoreset{equation}{section}
\renewcommand{\thefigure}{\hbox{\normalsize\arabic{section}.\arabic{figure}}}
\@addtoreset{figure}{section}
\renewcommand{\thetable}{\hbox{\normalsize\arabic{section}.\arabic{table}}}
\@addtoreset{table}{section}
\makeatother 

\usepackage{babel}
\makeatother
\begin{document}

\title{{\normalsize\begin{flushright}ITP Budapest Report No. 627\end{flushright}}\vspace{1cm}
Characterization of resonances using finite size effects}

\author{B. Pozsgay$^{1}$ and G. Takács$^{2}$\\
\\
$^{1}$\emph{Eötvös University, Budapest}\\
\emph{}\\
$^{2}$\emph{HAS Research Group for Theoretical Physics}\\
\emph{H-1117 Budapest, Pázmány Péter sétány 1/A}}

\date{13th April 2006}

\maketitle
\begin{abstract}
We develop methods to extract resonance widths from finite volume
spectra of 1+1 dimensional quantum field theories. Our two methods
are based on Lüscher's description of finite size corrections, and
are dubbed the Breit-Wigner and the improved {}``mini-Hamiltonian''
method, respectively. We establish a consistent framework for the
finite volume description of sufficiently narrow resonances that takes
into account the finite size corrections and mass shifts properly.
Using predictions from form factor perturbation theory, we test the
two methods against finite size data from truncated conformal space
approach, and find excellent agreement which confirms both the theoretical
framework and the numerical validity of the methods. Although our
investigation is carried out in 1+1 dimensions, the extension to physical
(3+1) space-time dimensions appears straightforward, given sufficiently
accurate finite volume spectra.
\end{abstract}

\section{\label{sec:Introduction} Introduction}

Two-dimensional field theories have attracted quite a lot of interest
for decades, partly because they are considered to be theoretical
laboratories for the development and testing of quantum field theory
methods. The present paper can be considered as one more step along
this line, taking up the issue of developing methods to extract resonance
parameters from finite volume spectra.

The framework of two-dimensional field theory is particularly well
adapted to this problem because non-integrable cases are increasingly
better understood. In integrable theories there can exist stable excitations
whose decay into lower mass particles would in principle be allowed
by their mass and conserved internal charges, but is prevented by
the infinitely many integrals of motion underlying integrability.
However, by adding a perturbation that breaks integrability one can
obtain a much richer phenomenology. At the same time, the underlying
integrable model provides us with very powerful tools: firstly, the
exact spectrum, scattering theory and form factors at the integrable
point are known, and secondly, form factor perturbation theory (FFPT)
\cite{nonintegrable,resonances} can give a prediction for the mass
shifts, phase shift corrections and decay widths when the non-integrable
perturbation is switched on. Therefore the phenomenology is under
firm control, which makes these theories into a very convenient testing
ground%
\footnote{It is also possible for resonances to occur in an integrable field
theory, as e.g. in the homogeneous sine-Gordon models \cite{hsg},
but the finite volume spectra of these models are not sufficiently
understood for the purposes of the present investigation. In addition,
the framework of perturbed integrable field theories, in which the
decay can be switched on with the integrability breaking perturbation,
provides us with a very useful control parameter, i.e. the perturbing
coupling, which can be used to tune the resonance width.%
}.

On the other hand, there exists a rather remarkable tool to compute
finite size spectra in two space-time dimensions, known as truncated
conformal space approach (TCSA), originally developed in \cite{yurov_zamolodchikov}
and carried further in many papers, in particular \cite{lassig_mussardo,ktw,SGTCSA}. 

Recently, Delfino et al. \cite{resonances} gave a detailed description
of the FFPT framework for the decay widths and performed calculations
for the Ising model. They also proposed a way to extract the decay
width from the finite size spectrum. However, the theoretical relation
to the finite volume spectrum was determined incorrectly as we show
in section \ref{sub:DGM_mismatch}. In addition, they could not reach
the required precision to extract any quantitatively useful information
from the TCSA spectrum either. In the course of our investigation we also
 present an improved realization 
of their proposal which is consistent with a Breit-Wigner resonance
description, and demonstrate its quantitative validity using
a more advanced TCSA analysis. In addition, we extend our considerations
to another prominent example of non-integrable field theory, the double
(two-frequency) sine-Gordon model.

We set out to formulate and test methods to extract (partial) decay
widths of unstable particles (resonances) from finite volume spectra,
using the ideas in \cite{luscher_resonance} as a starting point.
In particular we develop new and efficient methods to extract the
tiny effects of a narrow resonance with high precision, and test them
in the framework provided by two-dimensional quantum field theories.
We examine the scaling Ising model with a thermal and magnetic perturbation
(i.e. a continuum version of the familiar two-dimensional Ising lattice
away from critical temperature and in a magnetic field) and the double
(two-frequency) sine-Gordon model, since these are the most studied
examples of non-integrable quantum field theories in two dimensions
(see the references given in sections 2.2 and 2.3). However, the methods
we develop to extract the decay widths are very general, and their
higher-dimensional extension seems to be rather straightforward.

The paper is organized as follows. Section \ref{sec:ffpt} introduces
the framework of form factor perturbation theory and gives the theoretical
predictions for the two models considered. Section \ref{sec:resonance_finvol}
is devoted to the theoretical development of the methods used to extract
the decay widths from finite volume spectra, and examines the consistency
between them. Section \ref{sec:Numerical-methods} is devoted to the
specification of numerical methods and extrapolation procedures used,
and contains a discussion of the sources of numerical errors. We present
and discuss our numerical results in section \ref{sec:Results}, and
conclude briefly in section \ref{sec:Conclusions}. Two technical
details are relegated to separate appendices: one is the FFPT derivation
of the relevant decay width in the double sine-Gordon model and the
other is the detailed calculation performed to examine the consistency
of the methods developed in section \ref{sec:resonance_finvol}.

\section{\label{sec:ffpt} Unstable particles and form factor perturbation
theory }

\subsection{\label{sub:ffpt_general} General formalism}

In the framework of conformal perturbation theory, we consider a model
with the action \begin{equation}
\mathcal{A}(\mu,\lambda)=\mathcal{A}_{\mathrm{CFT}}-\mu\int dtdx\Phi(t,x)-\lambda\int dtdx\Psi(t,x)\label{eq:nonint_Lagrangian}\end{equation}
such that in the absence of the coupling $\lambda$, the model defined
by the action $\mathcal{A}(\mu,\lambda=0)$ is integrable. The two
perturbing fields are taken as scaling fields of the ultraviolet limiting
conformal field theory, with left/right conformal weights $h_{\Phi}=\bar{h}_{\Phi}<1$
and $h_{\Psi}=\bar{h}_{\Psi}<1$, i.e. they are relevant and have
zero conformal spin, resulting in a Lorentz-invariant field theory. 

The integrable limit $\mathcal{A}(\mu,\lambda=0)$ is supposed to
define a massive spectrum, with the scale set by the dimensionful
coupling $\mu$. The exact spectrum in this case consists of some
massive particles, forming a factorized scattering theory with known
$S$ matrix amplitudes, and characterized by a mass scale $M$ (which
we take as the mass of the fundamental particle generating the bootstrap),
which is related to the coupling $\mu$ via the mass gap relation\[
\mu=\kappa M^{2-2h_{\Phi}}\]
where $\kappa$ is a (non-perturbative) dimensionless constant. 

Switching on a second independent coupling $\lambda$ in general spoils
integrability, deforms the mass spectrum and the $S$ matrix, and
in particular allows decay of the particles which are stable at the
integrable point. One way to approach the dynamics of the model is
form factor perturbation theory initiated in \cite{nonintegrable}.
Let us denote the asymptotic states of the $\lambda=0$ theory by
\[
|A_{i_{1}}\left(\vartheta_{1}\right)\dots A_{i_{n}}\left(\vartheta_{n}\right)\rangle_{\lambda=0}\]
which describe \emph{in} states if the rapidities are ordered as $\vartheta_{1}>\dots>\vartheta_{n}$
and \emph{out} states for $\vartheta_{1}<\dots<\vartheta_{n}$. Then
the essential input to form factor perturbation theory consists of
the matrix elements of the local field $\Psi$ (so-called form factors):\[
F_{i_{1}\dots i_{n}}^{\Psi}\left(\vartheta_{1},\dots,\vartheta_{n}\right)=\langle0|\Psi(0,0)|A_{i_{1}}\left(\vartheta_{1}\right)\dots A_{i_{n}}\left(\vartheta_{n}\right)\rangle_{\lambda=0}\]
which (due to integrability at $\lambda=0$) can be obtained in a
closed exact form solving the form factor bootstrap axioms and identifying
the appropriate solution using known properties of the operator $\Psi$
(for a review of the form factor bootstrap, see \cite{smirnov} and
references therein). Given these data the following quantities can
be calculated to first order in $\lambda$:

\begin{enumerate}
\item The vacuum energy density is shifted by an amount\begin{equation}
\delta\mathcal{E}_{vac}=\lambda\left\langle 0\right|\Psi\left|0\right\rangle _{\lambda=0}.\label{vac_energy_shift}\end{equation}

\item The mass (squared) matrix $M_{ab}^{2}$ gets a correction\begin{equation}
\delta M_{ab}^{2}=2\lambda F_{a\bar{b}}^{\Psi}\left(i\pi\,,\,0\right)\delta_{m_{a},m_{b}}\label{mass_correction}\end{equation}
(where the bar denotes the antiparticle) supposing that the original
mass matrix was diagonal and of the form $M_{ab}^{2}=m_{a}^{2}\delta_{ab}\:.$
\item The scattering amplitude for the four particle process $a+b\,\rightarrow\, c+d$
is modified by \begin{equation}
\delta S_{ab}^{cd}\left(\vartheta,\lambda\right)=-i\lambda\frac{F_{\bar{c}\bar{d}ab}^{\Psi}\left(i\pi,\,\vartheta+i\pi,\,0,\,\vartheta\right)}{m_{a}m_{b}\sinh\vartheta}\quad,\quad\vartheta=\vartheta_{a}-\vartheta_{b}\:.\label{smatr_corr}\end{equation}
It is very important to keep in mind that this gives the variation
of the scattering phase when the center-off-mass energy (or, the Mandelstam
variable $s$) is kept fixed \cite{nonintegrable}. Therefore, in
terms of rapidity variables, this variation corresponds to the following:\[
\delta S_{ab}^{cd}\left(\vartheta,\lambda\right)=\frac{\partial S_{ab}^{cd}\left(\vartheta,\lambda=0\right)}{\partial\vartheta}\delta\vartheta+\lambda\left.\frac{\partial S_{ab}^{cd}\left(\vartheta,\lambda\right)}{\partial\lambda}\right|_{\lambda=0}\]
where \[
\delta\vartheta=-\frac{m_{a}\delta m_{a}+m_{a}\delta m_{a}+(m_{b}\delta m_{a}+m_{a}\delta m_{b})\cosh\vartheta}{m_{a}m_{b}\sinh\vartheta}\]
 is the shift of the rapidity variable induced by the mass corrections
(\ref{mass_correction}).
\end{enumerate}
It is not yet clear how to extend these results to second order. However,
it is possible to calculate the (partial) decay width of particles
\cite{resonances}. Suppose that the decay of particle $A_{c}$ into
particles $A_{a}$ and $A_{b}$ is kinematically allowed: $m_{c}>m_{a}+m_{b}$.
Then in the rest frame of particle $A_{c}$ the rapidities of the
outgoing particles are uniquely determined from energy-momentum conservation\begin{eqnarray*}
m_{a}\sinh\vartheta_{a}^{(cab)}+m_{b}\sinh\vartheta_{b}^{(cab)} & = & 0\\
m_{a}\cosh\vartheta_{a}^{(cab)}+m_{b}\cosh\vartheta_{b}^{(cab)} & = & m_{c}\end{eqnarray*}
and the partial decay width can be calculated as \begin{equation}
\Gamma_{c\rightarrow ab}=\lambda^{2}2^{1-\delta_{ab}}\frac{|F_{cab}^{\Psi}\left(i\pi,\vartheta_{a}^{(cab)},\vartheta_{b}^{(cab)}\right)|^{2}}{m_{c}^{2}m_{a}\left|\sinh\vartheta_{a}^{(cab)}\right|}\label{eq:decay_width}\end{equation}
All the masses $m_{a}$ in the above formulae correspond to those
in the unperturbed ($\lambda=0$) theory.

\subsection{\label{sub:ffpt_ising} Thermal perturbation of Ising model with
magnetic field}

The action \begin{equation}
\mathcal{A}_{\mathrm{Ising}}\left(h,\tau\right)=\mathcal{A}_{\mathrm{c=\frac{1}{2}}}-h\int dtdxd\sigma(t,x)-\tau\int dtdx\epsilon(t,x)\label{eq:ising_action}\end{equation}
describes the two-dimensional scaling Ising model, where $\mathcal{A}_{\mathrm{c=\frac{1}{2}}}$
is the action of the $c=1/2$ conformal field theory (free massless
Majorana fermion), describing the conformal limit of the two-dimensional
Ising model, $\epsilon$ is the energy density operator (which is
a primary field of weight $\Delta_{\epsilon}=\bar{\Delta}_{\epsilon}=1/2$)
and $\sigma$ is the magnetization operator ($\Delta_{\sigma}=\bar{\Delta}_{\sigma}=1/16$).
The coupling $h$ corresponds to the applied external magnetic field
and $\tau$ parameterizes the deviation of the temperature from criticality%
\footnote{Note that for this model the notation for the couplings is changed
from $\mu,\lambda$ to $h,\tau$ in order to conform with the notations
used in \cite{resonances}. In addition, $\Phi$ is now $\sigma$
and $\Psi$ is $\epsilon$.%
}. This model is considered as a prototype non-integrable field theory
in \cite{nonintegrable} and is widely discussed in the literature
\cite{ising1,ising2,ising3,ising4}. There are two options to consider
it as a perturbed integrable field theory. One way is to take $h$
as the perturbing coupling: in this case the Majorana fermions of
the thermally perturbed Ising model can be demonstrated to go through
some sort of confinement (McCoy-Wu scenario \cite{mccoy-wu}), and
the decay widths of the high-lying {}``meson'' states were computed
to $O(h^{3})$ in \cite{ising_pert_h}. Here we take the opposite
route and consider $\tau$ as the coupling that breaks integrability.

For $\tau=0$ the spectrum and the exact $S$ matrix is described
by the famous $E_{8}$ factorized scattering theory \cite{e8}, which
contains eight particles $A_{i},\; i=1,\dots,8$ with mass ratios
given by \begin{eqnarray*}
m_{2} & = & 2m_{1}\cos\frac{\pi}{5}\\
m_{3} & = & 2m_{1}\cos\frac{\pi}{30}\\
m_{4} & = & 2m_{2}\cos\frac{7\pi}{30}\\
m_{5} & = & 2m_{2}\cos\frac{2\pi}{15}\\
m_{6} & = & 2m_{2}\cos\frac{\pi}{30}\\
m_{7} & = & 2m_{4}\cos\frac{\pi}{5}\\
m_{8} & = & 2m_{5}\cos\frac{\pi}{5}\end{eqnarray*}
and the mass gap relation is \cite{phonebook}\[
m_{1}=(4.40490857\dots)|h|^{8/15}\]
or\begin{equation}
h=\kappa_{h}m_{1}^{8/15}\qquad,\qquad\kappa_{h}=0.06203236\dots\label{eq:ising_massgap}\end{equation}
We also quote the scattering phase shift of two $A_{1}$ particles
for $\lambda=0$, which has the form \begin{equation}
S_{11}(\vartheta)=\left\{ \frac{1}{15}\right\} \left\{ \frac{1}{3}\right\} \left\{ \frac{2}{5}\right\} \quad,\quad\{ x\}=\frac{\sinh\vartheta+i\sin\pi x}{\sinh\vartheta-i\sin\pi x}\label{eq:s11_ising}\end{equation}
All the other amplitudes $S_{ab}$ are determined by the $S$ matrix
bootstrap \cite{e8}.

The non-integrable model (\ref{eq:ising_action}) has a single dimensionless
parameter, which we choose as \[
t=\frac{\tau}{|h|^{8/15}}\]
The form factors of the operator $\epsilon$ in the $E_{8}$ model
were first calculated in \cite{delfino_simonetti} and their determination
was carried further in \cite{resonances}. Using the exact form factor
solutions the following results were obtained in \cite{resonances}
for the mass shift of $A_{a}$ and the partial decay width associated
to $A_{c}\rightarrow A_{a}+A_{b}$:\begin{eqnarray*}
\delta m_{a}^{2} & = & 2\tau f_{aa}\\
\Gamma_{c\rightarrow a+b} & = & \tau^{2}2^{1-\delta_{ab}}\frac{f_{cab}^{2}}{m_{c}^{2}m_{a}\sinh\vartheta_{a}^{(cab)}}\end{eqnarray*}
where \begin{eqnarray*}
f_{11} & = & (-17.8933\dots)\langle\epsilon\rangle\\
f_{22} & = & (-24.9467\dots)\langle\epsilon\rangle\\
f_{33} & = & (-53.6799\dots)\langle\epsilon\rangle\\
f_{44} & = & (-49.3206\dots)\langle\epsilon\rangle\\
f_{411} & = & (36.73044\dots)\langle\epsilon\rangle\\
f_{511} & = & (19.16275\dots)\langle\epsilon\rangle\\
f_{512} & = & (11.2183\dots)\langle\epsilon\rangle\end{eqnarray*}
and the vacuum expectation value of the perturbing field $\epsilon$
for $\tau=0$ is given by \cite{vevs}\[
\langle\epsilon\rangle=\epsilon_{h}|h|^{8/15}\qquad,\qquad\epsilon_{h}=2.00314\dots\]
The dimensionless decay width for the processes $A_{c}\rightarrow A_{1}+A_{1}$,
$c=4,5$ can be written as\begin{equation}
\frac{\Gamma_{c\rightarrow11}}{m_{1}}=t^{2}\frac{\left(\epsilon_{h}\kappa^{16/15}f'_{c11}\right)^{2}}{\left(\frac{m_{c}}{m_{1}}\right)^{2}\sqrt{\left(\frac{m_{c}}{2m_{1}}\right)^{2}-1}}\quad,\quad c=4,5\label{eq:isingdecaywidth}\end{equation}
where\begin{eqnarray*}
f'_{411} & = & 36.73044\dots\\
f'_{511} & = & 19.16275\dots\end{eqnarray*}

\subsection{\label{sub:ffpt_dsg} Double (two-frequency) sine-Gordon model}

Double sine-Gordon theory is another prototype of non-integrable field
theories which can be understood by application of techniques developed
in the context of integrable models \cite{delfino_mussardo}. It has
several interesting applications e.g. to the study of massive Schwinger
model (two-dimensional quantum electrodynamics), and in the description
of a generalized Ashkin-Teller model (a quantum spin system), both
of which are discussed in \cite{delfino_mussardo}. Another application
to the one-dimensional Hubbard model is examined in \cite{nersesyan}
(together with the generalized Ashkin-Teller model mentioned above).
A further potentially interesting application of the two-(and multi-)frequency
sine-Gordon model is the description of ultra-short optical pulses
propagating in resonant degenerate medium \cite{bullough}. Its phase
diagram was studied using non-perturbative finite size techniques
in \cite{dsg}, and the results were recently extended to the multi-frequency
generalization in \cite{tgzs}. There has been a certain doubt concerning
the validity of factor perturbation theory (or rather the proper way
of performing it) after Mussardo et al. \cite{semicl} applied a semiclassical
soliton form factor technique developed by Goldstone and Jackiw \cite{goldstone-jackiw},
and obtained results that contradict explicitely some of the results
derived from standard form factor perturbation theory in \cite{dsg}.
However, the conclusions drawn from the semiclassical technique were
shown to be untenable by extensive numerical work in \cite{dsg_mass},
which upheld the results of form factor perturbation theory as applied
in \cite{dsg}, and therefore we follow the same approach in this
paper.

The action of the two-frequency sine-Gordon model (with frequency
ratio $1:2$) is\begin{eqnarray}
 &  & \mathcal{A}=\mathcal{A}_{\mathrm{c=1}}+\mu\int dtdx\cos\beta\varphi+\lambda\int dtdx\cos\left(\frac{\beta}{2}\varphi+\delta\right)\label{eq:dsg_action}\\
 &  & \mathcal{A}_{\mathrm{c=1}}=\int dtdx\,\frac{1}{2}\left(\partial\varphi\right)^{2}\nonumber \end{eqnarray}
The spectrum at $\lambda=0$ consists of a soliton doublet with mass
$M$ related to $\mu$ via \cite{mass_scale}\begin{eqnarray}
\mu & = & \kappa_{\mathrm{sG}}(\xi)M^{2/(\xi+1)}\nonumber \\
\kappa_{\mathrm{sG}}(\xi) & = & \frac{2\Gamma(\frac{\xi}{1+\xi})}{\pi\Gamma(\frac{\xi}{1+\xi})}\left(\frac{\sqrt{\pi}}{2\Gamma\left(\frac{\xi+1}{2\xi}\right)\Gamma\left(\frac{\xi}{2}\right)}\right)^{2/(\xi+1)}\qquad,\qquad\xi=\frac{\beta^{2}}{8\pi-\beta^{2}}\label{eq:sg_massgap}\end{eqnarray}
and breathers $B_{n}$ with masses\begin{eqnarray}
m_{n} & = & 2M\sin\frac{n\pi\xi}{2}\qquad,\qquad n=1,\dots,\left[1/\xi\right]\label{eq:breather_spectrum}\end{eqnarray}
The exact $S$ matrix of all these particles was derived from the
axioms of factorized scattering in \cite{zam2}.

The strength of the non-integrable perturbation can be characterized
using the dimensionless ratio\[
t=\frac{\lambda}{M^{\frac{4+3\xi}{2+2\xi}}}\]
where $M$ is the soliton mass at the integrable point $\lambda=0$.
From (\ref{eq:breather_spectrum}) \[
\frac{m_{3}}{m_{1}}=1+2\cos\pi\xi>2\quad\mathrm{if}\quad\xi<1/3\]
so whenever $B_{3}$ exists, its decay to 2 $B_{1}$-s is always kinematically
allowed (for $\lambda$ small enough) and the same holds for all the
$B_{n}$, $n\geq3$. However, since \[
\frac{m_{2}}{m_{1}}=2\cos\frac{\pi\xi}{2}<2\]
 $B_{2}$ cannot decay to a pair of $B_{1}$-s (and therefore cannot
decay at all) if the perturbation $\lambda$ and correspondingly the
mass shifts are small enough. The decay width corresponding to the
simplest process $B_{3}\rightarrow B_{1}+B_{1}$ is calculated in
Appendix A for $\delta=-\pi/2$. The dimensionless decay width can
be written as \begin{equation}
\frac{\Gamma_{3\rightarrow11}}{M}=t^{2}\frac{s_{311}^{2}}{\left(\frac{m_{3}}{M}\right)^{2}\frac{m_{1}}{M}\sqrt{\left(\frac{m_{3}}{2m_{1}}\right)^{2}-1}}\label{eq:sg_decaywidth}\end{equation}
where $s_{311}$ is given by (\ref{eq:s311_result}). Because the
$B_{3}$ state is the lowest one whose decay can be observed, it is
preferable to use in the comparison with truncated conformal space
to minimize the truncation errors. In addition, there is a $\mathbb{Z}_{2}$
symmetry at $\delta=-\pi/2$ which can be turned to a significant
numerical advantage (see the analysis in subsection \ref{sub:Truncated-conformal-space}
for details).

\section{\label{sec:resonance_finvol} Resonance widths from finite volume
spectra}

In this section we discuss two different approaches to linking the
finite volume spectra with the theoretical predictions of form factor
perturbation theory in the previous section. The first one uses a
Breit-Wigner parameterization for the finite volume levels in the
vicinity of the volume $L_{0}$ where the one particle level $A_{c}$
crosses the two-particle level $A_{a}A_{b}$ before switching on the
coupling. This method relies on obtaining the two particle phase shift
from the finite size spectrum using Lüscher's results \cite{luscher_2particle}
and then finding the Breit-Wigner resonance parameters directly from
the phase shift. The second approach uses a {}``mini-Hamiltonian''
description for the levels to extract directly the form factor determining
the decay width. The two approaches are then compared and we learn
how to improve the results of Delfino et al. \cite{resonances} for
the {}``mini-Hamiltonian'' to make them consistent with the Breit-Wigner
description (for a narrow resonance).

\subsection{\label{sub:decayrate_extract} Lüscher's formulation: methods to
extract the decay rate from the finite size spectrum}

Suppose that the finite volume spectrum of the theory is given in
a form of functions $E_{i}(L)$ describing the energy of the $i$th
excited stated as a function of volume $L$, normalized to the ground
state energy in the same volume. Suppose further that we investigate
a decay process $A_{c}\rightarrow A_{a}+A_{b}$ ($m_{c}>m_{a}+m_{b}$)
and that we have a parameter $\lambda$ describing the interaction
responsible for the decay, such that for $\lambda=0$ the particle
$A_{c}$ is stable (the notation here follows the conventions of section
\ref{sub:ffpt_general}).

We can then find the energy level corresponding asymptotically to
a stationary particle $A_{c}$ at $\lambda=0$, denoted by $E_{c}(L)$
and two-particle levels composed of $A_{a}$ and $A_{b}$ with zero
total momentum, denoted by $E_{ab}(L)$ (where we suppressed an index
labeling states corresponding to different relative momenta of $a$
and $b$). According to \cite{luscher_2particle}, up to corrections
vanishing exponentially with $L$ (which we neglect from now on),
the finite size correction to the two-particle level is determined
by the equations\begin{eqnarray}
m_{a}L\sinh\vartheta_{a}+\delta_{ab}\left(\vartheta_{a}-\vartheta_{b}\right) & = & 2n_{a}\pi\nonumber \\
m_{b}L\sinh\vartheta_{b}+\delta_{ab}\left(\vartheta_{b}-\vartheta_{a}\right) & = & 2n_{b}\pi\label{eq:Luscher_quant}\end{eqnarray}
and\[
E_{ab}(L)=m_{a}\cosh\vartheta_{a}+m_{b}\cosh\vartheta_{b}\]
where $\delta_{ab}\left(\vartheta\right)=-\delta_{ab}\left(-\vartheta\right)$
is the elastic two-particle phase shift defined from the elastic two-particle
scattering phase $S_{ab}$ via the relation\[
S_{ab}\left(\vartheta\right)=\mathrm{e}^{i\delta_{ab}\left(\vartheta\right)}\]
 and $n=n_{a}=-n_{b}$ for zero total momentum labels the different
possible values for the relative momentum. As a result, the energy
of any two particle level decreases as $1/L^{2}$, while the finite
size corrections of a one-particle level are known to vanish exponentially
fast with increasing volume \cite{luscher_1particle}. In the infinite
volume limit $E_{ab}(L)<E_{c}(L)$ (since in the limit $L\,\rightarrow\,\infty$
they tend to $m_{a}+m_{b}$ and $m_{c}$, respectively) and therefore
there is a volume $L_{0}$ where the two levels meet: $E_{ab}(L_{0})=E_{c}(L_{0})=E_{0}$.

\subsubsection{\label{sub:miniham_method} The {}``mini-Hamiltonian''}

Switching on a small $\lambda\neq0$ the levels only move a small
amount and thus can be easily re-identified, however the degeneracy
at $L_{0}$ is lifted by an amount related to the width of the decay
process (as we shall see shortly). Let us restrict ourselves to the
two-dimensional subspace spanned by the two levels. We suppose that
the relevant region in the volume $L$ is of order $\lambda$, which
turns out to be a self-consistent assumption. Then for $L\sim L_{0}$
and $\lambda\sim0$ the effective {}``mini-Hamiltonian'' in this
subspace which determines the behaviour of the levels to leading order
in $L-L_{0}$ and $\lambda$ can be parameterized as follows\begin{equation}
H=E_{0}+\left(L-L_{0}\right)\left(\begin{array}{cc}
\alpha_{1}\\
 & \alpha_{2}\end{array}\right)+\lambda\left(\begin{array}{cc}
A(L) & B(L)\\
B(L) & C(L)\end{array}\right)\label{eq:miniham}\end{equation}
where the relative phases of the two states were chosen such that
$B$ is real and nonnegative ($A$, $C$ are real by hermiticity).
According to the prescription of degenerate perturbation theory, the
splitting of the two levels is given by diagonalizing (\ref{eq:miniham})
in the subspace of the two levels: \[
\delta E(L)=\sqrt{\left((\alpha_{1}-\alpha_{2})(L-L_{0})+\lambda(A(L)-C(L))\right)^{2}+4B^{2}\lambda^{2}}\]
Neglecting the volume dependence of $A,\; B$ and $C$ for the moment,
the minimum splitting between the two levels occurs at \begin{equation}
L_{\mathrm{min}}-L_{0}=-\lambda\frac{A-C}{\alpha_{1}-\alpha_{2}}\label{eq:Lmin}\end{equation}
which shows that it was consistent to suppose that the relevant values
of $L-L_{0}$ are of the order $\lambda$. The minimal energy split
is then\begin{equation}
\delta E(L_{\mathrm{min}})=2|B\lambda|\label{eq:miniham_readout}\end{equation}
This shows that to first order in $\lambda$ it is consistent to replace
the coefficients $A(L),B(L),C(L)$ by their values taken at $L_{0}$.
This observation will be used several times in the sequel.

\begin{figure}
\begin{center}\includegraphics{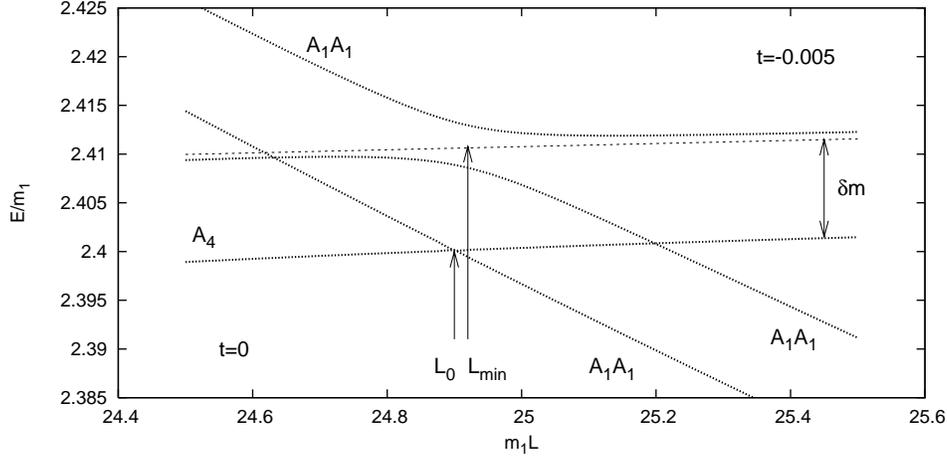}\end{center}

\caption{\label{cap:levels} Behaviour of levels at $\lambda=0$ and $\lambda\neq0$,
illustrated using actual numerical data for the Ising case. (Energy
and distance are measured in units given by the value that the mass
of the lightest particle ($m_{1}$) takes at $\lambda=0$.)}
\end{figure}

The behaviour of levels discussed above is illustrated in figure \ref{cap:levels}.

\subsubsection{\label{sub:Breit-Wigner-method} Breit-Wigner analysis}

\begin{figure}
\begin{center}\subfigure[$\lambda=0$]{\includegraphics[%
  scale=0.6,
  angle=270]{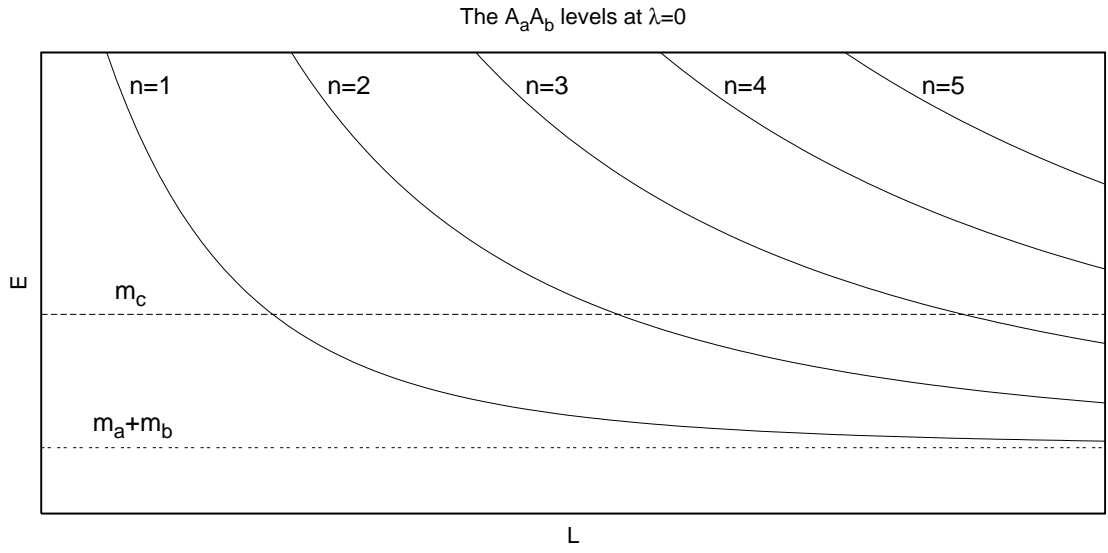}}\\
\subfigure[$\lambda\neq 0$]{\includegraphics[%
  scale=0.6,
  angle=270]{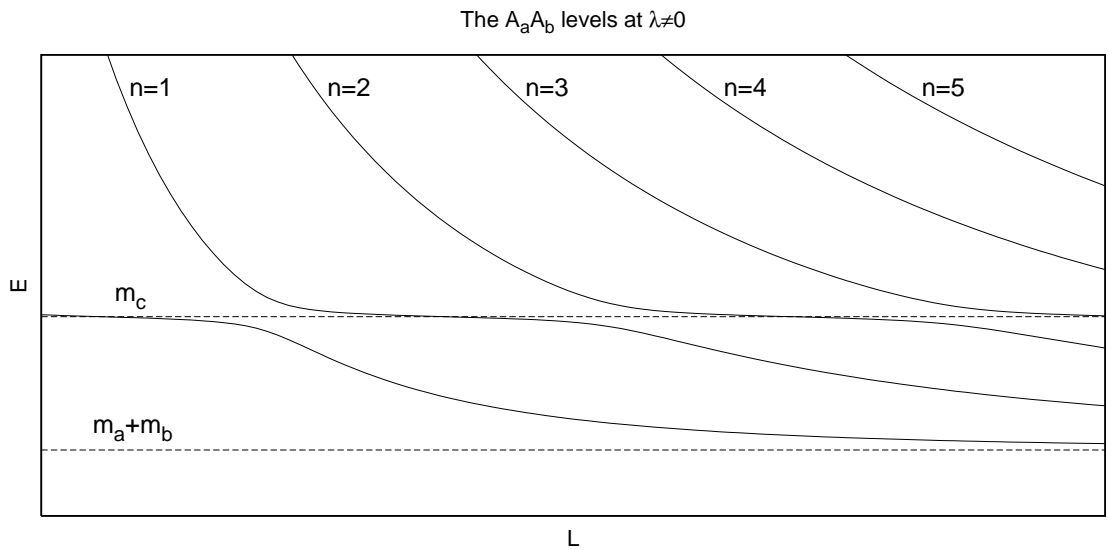}}\end{center}

\caption{\label{cap:two_part_state_illustration} Illustrating the behaviour
of two-particle states in the integrable case $\lambda=0$ and with
a resonance present ($\lambda\neq0$). }
\end{figure}
\begin{figure}
\begin{center}\includegraphics{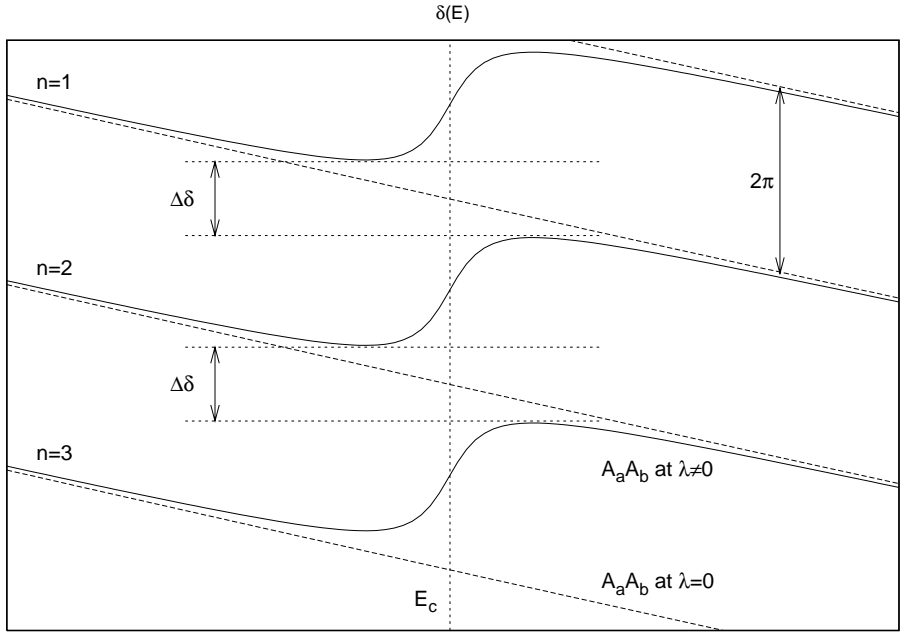}\end{center}

\caption{\label{cap:phase_illustration}Phase shift functions $\delta_{n}(E)$
extracted from the two-particle levels illustrated in figure \ref{cap:two_part_state_illustration}
around the resonant energy $E_{c}$ .}
\end{figure}

Here we build on the approach advocated by Lüscher in \cite{luscher_resonance},
but with some modification to make it more suited to numerical analysis.
At $\lambda\neq0$ there are no stable one-particle states of $A_{c}$
anymore, and therefore asymptotically all the states are identified
as two-particle states $A_{a}A_{b}$. Due to the lifting of the degeneracies
in the crossings with the one-particle level of $A_{c}$ these states
acquire plateaux as illustrated in figure \ref{cap:two_part_state_illustration},
which correspond to the appearance of a Breit-Wigner resonance contribution
in the two-particle phase shift: \[
\delta_{ab}(E)=\delta_{0}(E)+\delta_{BW}(E)\]
(as a function of the center-of-mass energy $E$), where $\delta_{0}$
is the background part of the phase shift, and the Breit-Wigner contribution
has the form \[
\delta_{BW}(E)=-i\log\frac{E-E_{c}-i\Gamma/2}{E-E_{c}+i\Gamma/2}\]
with $E_{c}$ as the center and $\Gamma$ as the width of the resonance.
If \[
\beta=\left.\frac{d\delta_{0}}{dE}\right|_{E=E_{c}}<0\]
then the total phase shift has two extrema (a minimum at $E<E_{c}$
and a maximum at $E>E_{c}$, see figure \ref{cap:phase_illustration}).
Their positions can be calculated to first order in $\Gamma$:\[
E_{\pm}=E_{c}\pm\sqrt{-\frac{\Gamma}{\beta}}\]
with the values\[
\delta_{ab}\left(E_{\pm}\right)=\pi+\delta_{0}(E_{c})\pm\left(\pi-2\sqrt{-\beta\Gamma}\right)\]
Now let us take two neighbouring two-particle levels $E_{1}(L)$and
$E_{2}(L)$. Using the description (\ref{eq:Luscher_quant}) we can
define the phase shift functions for the two levels as follows: first
solve\[
E=\sqrt{p^{2}+m_{a}^{2}}+\sqrt{p^{2}+m_{b}^{2}}\rightarrow p(E)\]
and then extract the phase shift as \begin{eqnarray}
\delta_{1}(E) & = & -L_{1}(E)p(E)\nonumber \\
\delta_{2}(E) & = & -L_{2}(E)p(E)\label{eq:BW_phaseshift12def}\end{eqnarray}
where $L_{1,2}(E)$ are the volume-energy functions for the two states
(here we supposed that the two-particle state has zero total momentum,
and the momenta of the particles are $p$ and $-p$). The functions
$\delta(E)$ are illustrated in figure \ref{cap:phase_illustration}.
Since\begin{eqnarray*}
\delta_{1}(E) & = & \delta_{ab}\left(E\right)-2n_{1}\pi\\
\delta_{2}(E) & = & \delta_{ab}\left(E\right)-2n_{2}\pi\end{eqnarray*}
where $n_{1}$ and $n_{2}$ are the momentum quantum numbers which
for neighbouring states satisfy $n_{2}=n_{1}+1$, we obtain \begin{equation}
\Delta\delta=\min\delta_{1}(E)-\max\delta_{2}(E)=4\sqrt{-\beta\Gamma}\label{eq:BW_readout}\end{equation}
This gives a method to extract the value of $\Gamma$ from the finite
volume spectrum: determine first the phase shift functions $\delta_{1,2}$
from two neighbouring two-particle levels and then take the difference
between their extrema to find $\sqrt{\Gamma}$. Since $\Gamma$ is
of order $\lambda^{2}$, it follows that $\sqrt{\Gamma}$ is of order
$\lambda$ which means that it is much easier to measure than $\Gamma$
in the small $\lambda$ regime when the resonance is narrow. This
is a huge advantage over previously proposed methods which gave $\Gamma$
directly, like the approach advocated in \cite{luscher_resonance}
which relates $\Gamma$ to the slope of the plateaux at their middle
point (more precisely at the value of the volume when the resonant
contribution to the phase shift $\delta$ of the given level passes
through $\pi$). Indeed it is already apparent from figure \ref{cap:two_part_state_illustration}
that the level splittings determining $\Delta$ are much easier to
observe compared to the slope at the middle of the plateaux. In addition,
$\Delta$ can be extracted by measuring the spectrum in a small neighbourhood
of $L_{\mathrm{min}}$ which makes the \emph{residual} finite size
effects (those that decay exponentially on the volume) much easier
to control.

\begin{figure}
\begin{center}\includegraphics{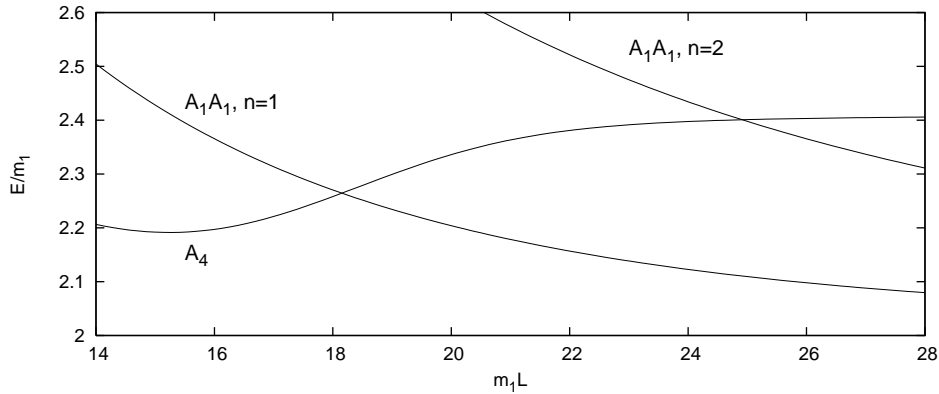}\end{center}

\caption{\label{cap:residual_fse}Illustration of the residual finite size
effects in the Ising model, for the levels pertaining to $A_{4}$
and two-particle states $A_{1}A_{1}$. The volume dependence of the
two-particle levels follows very precisely the description given in
eq. (\ref{eq:Luscher_quant}), but the variation of $A_{4}$ is entirely
due to the residual finite size effects that we neglected. The plot
shows the integrable case $t=0$ when the line crossings are exact,
but the only result of switching on a small value for $t$ is a slight
shift in the lines and resolution of the degeneracy at the level crossings. }
\end{figure}

Finally we remark that while the simplest possibility of extracting
the resonance parameters would seem to be the method of fitting the
numerically determined phase shift function $\delta(E)$ with a Breit-Wigner
resonance function, in practice this has the severe disadvantage that
the finite size data necessary to extract the relevant part of the
phase shift cover an extended range between two line crossings together
with a neighbourhood of the crossings themselves. Over such an extended
range, the contribution of residual finite size corrections varies
substantially, which causes such a significant distortion of the shape
of the Breit-Wigner resonance contribution that there is no way to
perform a reliable fit to the data. This problem is illustrated in
figure \ref{cap:residual_fse}, where it is obvious how much the one-particle
level varies between two adjacent level crossings. This also makes
the application of Lüscher's {}``plateau slope'' method impossible,
which can be seen from the fact that in the absence of residual finite
size effects we would expect the slope to be negative, as illustrated
in figure \ref{cap:two_part_state_illustration}. However, it is obvious
from the data in figure \ref{cap:residual_fse} that the leading residual
finite size corrections are such that the particle masses approach
their infinite volume from below, which is known to be a generic feature
in two dimensional field theories \cite{klassen_melzer}.

\subsubsection{\label{sub:miniham_BW_link} Linking the {}``mini-Hamiltonian''
with the Breit-Wigner formula}

We can now match this method with the previous one. For simplicity
we shall assume that the two decay products are identical to a particle
denoted by $A_{1}$ (i.e. $a=b=1$). Parameterizing the two levels
using (\ref{eq:miniham}) we can express the parameters in the {}``mini-Hamiltonian''
by extracting \[
\Delta\delta=\min\delta_{1}(E)-\max\delta_{2}(E)\]
from the levels and matching it with the result (\ref{eq:BW_readout})
of the Breit-Wigner analysis. From (\ref{eq:miniham}), the two levels
read\[
E_{1,2}=E_{0}+\frac{A+C}{2}\lambda+\frac{1}{2}\left(\alpha_{1}+\alpha_{2}\right)(L-L_{0})\pm\frac{1}{2}\sqrt{(\alpha_{1}-\alpha_{2})^{2}(L-L_{\mathrm{min}})^{2}+4B^{2}\lambda^{2}}\]
($E_{1}/E_{2}$ corresponds to the sign choice $+/-$, respectively)
where $L_{\mathrm{min}}$ is given by eq. (\ref{eq:Lmin}). At $\lambda=0$
\[
E_{1,2}=E_{0}+\frac{1}{2}\left(\alpha_{1}+\alpha_{2}\right)\left(L-L_{0}\right)\pm\frac{1}{2}\left(\alpha_{1}-\alpha_{2}\right)|L-L_{0}|\]
Recalling that $m_{c}$ is the mass of the particle $A_{c}$ whose
decay process we are interested in, we get $E_{0}=m_{c}$ (where we
neglected the residual finite size effects decaying exponentially
with the volume, as before). Furthermore, it is known that $E_{2}=m_{c}$
for $L<L_{0}$ and $E_{1}=m_{c}$ at $L>L_{0}$ (due to the level
crossing at $L_{0}$, the interpretation of the levels in terms of
asymptotic particle states swaps over). As a result, $\alpha_{1}=0$
and $\alpha_{2}=-2\alpha<0$ , and so the expression for the two levels
turns into\begin{equation}
E_{1,2}=m_{c}+A\lambda-\alpha(L-L_{\mathrm{min}})\pm\sqrt{\alpha^{2}(L-L_{\mathrm{min}})^{2}+B^{2}\lambda^{2}}\label{eq:miniham_levels12}\end{equation}
with\begin{equation}
L_{\mathrm{min}}=L_{0}-\lambda\,\frac{A-C}{2\alpha}\label{eq:Lminnew}\end{equation}
and denoting \[
A=A(L_{0})\quad,\quad B=B(L_{0})\quad,\quad C=C(L_{0})\]
since as pointed out in section \ref{sub:miniham_method}, the coefficients
$A(L),B(L),C(L)$ can be substituted with their values taken at $L=L_{0}$
if one calculates only to the lowest order in $\lambda$.

Note that for $|\alpha(L-L_{\mathrm{min}})|\gg|B\lambda|$ one of
the levels (the identity of which depends on the sign of $L-L_{\mathrm{min}}$)
is $L$-independent and takes the value\begin{equation}
E_{c}\left(\lambda\right)=m_{c}+A\lambda\label{eq:miniham_masshift}\end{equation}
which therefore can be identified with the mass of the resonance corresponding
to the unstable particle $A_{c}$, to first order in $\lambda$ (this
shift is also illustrated in figure \ref{cap:levels}). 

Using form factor perturbation theory (\ref{mass_correction}) we
obtain the identification\[
A=\frac{f_{cc}}{m_{c}}\qquad,\qquad f_{cc}=F_{cc}(i\pi,0).\]

After some calculation (the details are given in Appendix B) the phase
shift difference can also be extracted\[
\min\delta_{1}-\max\delta_{2}=2B\lambda\sqrt{-\frac{1}{\alpha}\sqrt{m_{c}^{2}-4m_{1}^{2}}\left.\frac{d\delta_{0}}{dE}\right|_{E=E_{0}}}\]
where $\delta_{1,2}$ are the phase shift functions for the levels
$E_{1,2}(L)$ as defined in (\ref{eq:BW_phaseshift12def}). Now this
must be compared to the result from the Breit-Wigner analysis (cf.
eq. (\ref{eq:BW_readout}) )\[
\min\delta_{1}-\max\delta_{2}=4\sqrt{-\Gamma\left.\frac{d\delta_{0}}{dE}\right|_{E=E_{0}}}\]
This way we obtain the relation\[
\Gamma=\frac{B^{2}\lambda^{2}}{4\alpha}\sqrt{m_{c}^{2}-4m_{1}^{2}}\]
On the other hand, using the form factor perturbation theory result
(\ref{eq:decay_width}) gives \begin{eqnarray*}
\Gamma=\lambda^{2}\frac{f_{c11}^{2}}{m_{c}^{2}m_{1}\left|\sinh\vartheta_{1}^{(c11)}\right|}\quad,\quad &  & f_{c11}=\left|F_{c11}^{\Psi}\left(i\pi,\vartheta_{1}^{(c11)},\vartheta_{1}^{(c11)}\right)\right|\\
 &  & m_{c}=2m_{1}\cosh\vartheta_{1}^{(c11)}\end{eqnarray*}
and so \[
B=\frac{f_{c11}}{m_{c}}\sqrt{\frac{8\alpha}{m_{c}^{2}-4m_{1}^{2}}}\]
We also need an expression for $\alpha$. At $\lambda=0$ we can express
it using the energy $E_{11}(L)$ of the $A_{1}A_{1}$ two-particle
state from (\ref{eq:tplevel_expansion})\[
\alpha=\left.-\frac{1}{2}\frac{dE_{11}}{dL}\right|_{L=L_{0}}\]
Writing the energy of the two-particle level with the one-particle
momentum $p$ as $E_{11}(L)=2\sqrt{p(L)^{2}+m_{1}^{2}}$ and using
$E_{11}(L_{0})=m_{c}+O(\lambda)$ we get\begin{equation}
B=\frac{2f_{c11}}{m_{c}^{3/2}}\sqrt{\left.\left(-\frac{1}{p}\frac{dp(L)}{dL}\right)\right|_{L=L_{0}}}\label{eq:Bminiham_BW}\end{equation}
Putting $B=0$ the level corresponding to the two-particle state $A_{1}A_{1}$
takes the following value at $L_{0}$:\[
E_{11}\left(L_{0}\right)=m_{c}+C\lambda\]
so $C\lambda$ is the energy shift of the two-particle level at the
volume $L=L_{0}$ if the resonance contribution is neglected. This
is also consistent with eq. (\ref{eq:Lminnew}) which for $B=0$ gives
the location of the level crossing to first order in $\lambda$. This
observation provides a possibility to determine $C$ (for details
see Appendix \ref{sub:C_yangbethe}), with the result\begin{equation}
C=\left.\left(-\frac{1}{p}\frac{dp(L)}{dL}\right)\right|_{L=L_{0}}\left(\frac{4f_{1111}}{m_{c}^{2}}+L_{0}\frac{4f_{11}}{m_{c}}\right)\label{eq:C_yangbethe_result}\end{equation}

\subsection{\label{sub:miniham_theory} Theoretical determination of the {}``mini-Hamiltonian''}

\subsubsection{\label{sub:DGM_mismatch} Mismatch between the Breit-Wigner formalism
and the DGM prediction}

Now let us turn to theoretical determination of the {}``mini-Hamiltonian''
(\ref{eq:miniham})

\[
H=E_{0}+\left(L-L_{0}\right)\left(\begin{array}{cc}
\alpha_{1}\\
 & \alpha_{2}\end{array}\right)+\lambda\left(\begin{array}{cc}
A(L) & B(L)\\
B(L) & C(L)\end{array}\right)\]
 It was shown in subsection \ref{sub:miniham_BW_link} that matching
the {}``mini-Hamiltonian description'' with the Breit-Wigner analysis
gives the relations\begin{equation}
A_{\mathrm{BW}}(L_{0})=\frac{f_{cc}}{m_{c}}\quad,\quad B_{\mathrm{BW}}(L_{0})=\frac{2f_{c11}}{m_{c}^{3/2}}\sqrt{\left.\left(-\frac{1}{p}\frac{dp(L)}{dL}\right)\right|_{L=L_{0}}}\label{eq:miniham_BW}\end{equation}
In addition, matching the Lüscher description (\ref{eq:Luscher_quant})
of the two-particle states with the {}``mini-Hamiltonian'' also
gave (\ref{eq:C_yangbethe_result}) for $C$.

Delfino, Grinza and Mussardo (DGM for short) obtain the following
results for the matrix elements \cite{resonances}\begin{equation}
A_{\mathrm{DGM}}(L)=\frac{f_{cc}}{m_{c}}\quad,\quad B_{\mathrm{DGM}}(L)=\frac{2f_{c11}}{\sqrt{m_{c}^{3}L}}\quad,\quad C_{\mathrm{DGM}}(L)=\frac{4f_{1111}}{m_{c}^{2}L}\label{eq:miniham_DGM}\end{equation}
where \begin{equation}
f_{cc}=F_{c\bar{c}}^{\Psi}\left(i\pi\,,\,0\right)\quad,\quad f_{1111}=\lim_{\delta\rightarrow0}F_{1111}^{\Psi}\left(\vartheta_{1}^{(c11)}+i\pi-\delta,-\vartheta_{1}^{(c11)}+i\pi-\delta,-\vartheta_{1}^{(c11)},\vartheta_{1}^{(c11)}\right)\label{eq:fcc_f1111_def}\end{equation}
and\begin{equation}
f_{c11}=\left|F_{c11}^{\Psi}\left(i\pi,\vartheta_{1}^{(c11)},\vartheta_{1}^{(c11)}\right)\right|\label{eq:fc11}\end{equation}
(recall that the phase of the matrix element $B(L)$ can be transformed
away by redefining the relative phase of the two states, so we can
take the absolute value in (\ref{eq:fc11})).

We can see that the result for $A(L_{0})$ matches perfectly, but
$B(L_{0})$ agrees only if\[
\left.\left(-\frac{1}{p}\frac{dp(L)}{dL}\right)\right|_{L=L_{0}}=\frac{1}{L_{0}}\]
This is only true if we neglect the background phase shift $\delta_{0}$
since the quantization rule for the momentum is\begin{eqnarray*}
pL+\delta_{0}\left(E\right) & = & 2n\pi\\
E & = & 2\sqrt{p^{2}+m_{1}^{2}}\end{eqnarray*}
Putting $\delta_{0}=0$\[
p(L)=\frac{2n\pi}{L}\quad\Rightarrow\quad\left.\left(-\frac{1}{p}\frac{dp(L)}{dL}\right)\right|_{L=L_{0}}=\frac{1}{L_{0}}\]
However, the background phase shift is already present at $\lambda=0$
and therefore the formulae (\ref{eq:miniham_DGM}) are not consistent
with the Breit-Wigner analysis to first order in $\lambda$. In fact,
the background phase shift (especially the fact that $\delta_{0}'(E_{0})<0$)
played an important role in the considerations based on the Breit-Wigner
resonance parameterization. 

In addition, the DGM prediction for $C$ does not agree with (\ref{eq:C_yangbethe_result})
and so is inconsistent with the behaviour of two-particle states in
finite volume even in the absence of a resonance. Taken together,
this means that the DGM formulae (\ref{eq:miniham_DGM}) cannot be
used self-consistently to extract the form factors from the behaviour
of the levels for small $\lambda$.

\subsubsection{\label{sub:improved_miniham} Rederiving the {}``mini-Hamiltonian''
from the Lagrangian approach}

We shall now resolve the above problem by reexamining the derivation
of the {}``mini-Hamiltonian'' from the Lagrangian approach. We normalize
the one-particle states in infinite volume as \[
\langle A_{a}(p_{a})|A_{b}(p_{b})\rangle=2\pi\delta_{ab}\delta(p_{a}-p_{b})E_{a}\]
In finite volume the one-particle momenta (up to corrections vanishing
exponentially with the volume) are quantized as \[
p_{a}=\frac{2\pi n_{a}}{L}\]
with the density of states \[
\frac{dn}{dp}=\frac{L}{2\pi}\]
therefore the normalization reads \[
\langle A_{a}(p_{a})|A_{b}(p_{b})\rangle=\delta_{ab}\delta_{p_{a}p_{b}}LE_{a}\]
To obtain an orthonormal basis the one-particle states in finite volume
must be related to the ones in infinite volume as\[
|A_{a}(p_{a})\rangle_{L}=\frac{1}{\sqrt{E_{a}L}}|A_{a}(p_{a})\rangle\]
The normalization of two-particle states in infinite volume reads\[
\langle A_{a}(p_{a})A_{b}(p_{b})|A_{c}(p_{c})A_{d}(p_{d})\rangle=2\pi\delta_{ac}\delta(p_{a}-p_{c})E_{a}2\pi\delta_{bd}\delta(p_{b}-p_{d})E_{b}\]
Taking the density of the two-particle states to be the product of
the one-particle densities, the following expression is obtained for
the states normalized in finite volume $L$\[
|A_{a}(p_{a})A_{b}(p_{b})\rangle_{L}=\frac{1}{\sqrt{E_{a}L}}\frac{1}{\sqrt{E_{b}L}}|A_{a}(p_{a})A_{b}(p_{b})\rangle\]
In particular, considering states containing two identical particles
and having zero total momentum we get\[
|A_{a}(p_{a})A_{a}(-p_{a})\rangle_{L}=\frac{2}{EL}|A_{a}(p_{a})A_{a}(-p_{a})\rangle\]
where $E=2E_{a}$ is the total energy of the state. The perturbing
operator reads \[
H'=\lambda\int_{0}^{L}dx\Psi(x)\]
The mini-Hamiltonian can be computed taking the matrix elements of
$H'$ in the basis \[
|A_{c}(0)\rangle_{L=L_{0}}\qquad,\qquad|A_{1}(p)A_{1}(-p)\rangle_{L=L_{0}}\]
Using translational invariance to perform the spatial integration,
and the fact that the energy of these states is $E_{c}(L_{0})=E_{11}(L_{0})=m_{c}$
the expression of the {}``mini-Hamiltonian'' in terms of the form
factors in infinite volume becomes\begin{eqnarray}
A(L_{0}) & = & L_{0}\frac{1}{m_{c}L_{0}}\langle A_{c}(0)|\Psi(0)|A_{c}(0)\rangle=\frac{f_{cc}}{m_{c}}\nonumber \\
B(L_{0}) & = & L_{0}\frac{2}{(m_{c}L_{0})^{3/2}}\langle A_{c}(0)|\Psi(0)|A_{1}(p)A_{1}(-p)\rangle=\frac{2f_{c11}}{\sqrt{m_{c}^{3}L}_{0}}\nonumber \\
C(L_{0}) & = & L_{0}\frac{4}{(m_{c}L_{0})^{2}}\langle A_{1}(p)A_{1}(-p)|\Psi(0)|A_{1}(p)A_{1}(-p)\rangle=\frac{4f_{1111}}{m_{c}^{2}L_{0}}+\frac{4f_{11}}{m_{c}}\label{eq:miniham_naive}\end{eqnarray}
where \[
f_{11}=F_{11}^{\Psi}\left(i\pi,0\right)\]
$A(L_{0})$ and $B(L_{0})$ does agree with the DGM result (\ref{eq:miniham_DGM}),
but $C(L_{0})$ contains a new term coming from the disconnected part
of the four-particle matrix element. The origin of this contribution
is that in contrast to the bulk setting discussed in \cite{nonintegrable},
the finite volume Hamiltonian does not include any counter terms for
the mass shifts (for details see Appendix \ref{sub:C_disconnected}).
The presence of the disconnected term is tested using TCSA numerical
data in subsection \ref{sub:ising_C_test}.

However, we already know that even (\ref{eq:miniham_naive}) is inconsistent
with the proper finite size description, which gives (\ref{eq:C_yangbethe_result},\ref{eq:miniham_BW})
to leading order in $\lambda$. The underlying reason is that \emph{the
density of two-particle states is not a product of the densities of
one-particle} \emph{states}: the presence of the background phase
shift modifies it by terms of order $L^{-1}$ already at order $\lambda^{0}$.
As previously, we continue to neglect residual finite size corrections
(i.e. those that decrease exponentially with the volume), but all
power-like corrections must be taken into account (at least to the
necessary order in $\lambda$) in order to maintain consistency. 

To determine the correct normalization of the states $|A_{1}(p_{1})A_{1}(p_{2})\rangle$
let us introduce the total momentum $P=p_{1}+p_{2}$ and the relative
momentum $p=(p_{1}-p_{2})/2$. The Jacobian of this change of variables
is $1$ and so (in infinite volume)\[
\langle A_{1}(p'_{1})A_{1}(p'_{2})|A_{1}(p_{1})A_{1}(p_{2})\rangle=(2\pi)^{2}\delta(P-P')\delta(p-p')\sqrt{p_{1}^{2}+m_{1}^{2}}\sqrt{p_{2}^{2}+m_{1}^{2}}\]
In finite volume, the density of states in the relative momentum variable
can be obtained from the quantization condition (written here for
the case of zero momentum ie. $p_{1}=-p_{2}=p$) \begin{eqnarray}
pL+\delta(E(p)) & = & 2n\pi\nonumber \\
E(p) & = & 2\sqrt{p^{2}+m_{1}^{2}}\label{eq:quant_cond_density}\end{eqnarray}
The density of states in the total momentum variable is unaffected,
since the phase shift drops from the quantization of the total momentum,
so it is only the density in the relative momentum of the two particles
which must be modified. Let us examine the equation \[
pL+\delta(E(p))=2n\pi\]
Taking the derivative of this equation with respect to $p$, first
at fixed $L$ and then at fixed $n$:\begin{eqnarray*}
L+\frac{d\delta(E(p))}{dp} & = & 2\pi\left.\frac{\partial n}{\partial p}\right|_{L}\\
L+\frac{d\delta(E(p))}{dp} & = & -p\left.\frac{\partial L}{\partial p}\right|_{n}\end{eqnarray*}
We obtain that the density of states at can be expressed as\[
\frac{dn}{dp}=-\frac{1}{2\pi}p\frac{dL(p)}{dp}\]
where the derivative \[
\frac{dL(p)}{dp}=\left.\frac{\partial L}{\partial p}\right|_{n}\]
 is exactly the inverse of\[
\frac{dp(L)}{dL}\]
 that was encountered in (\ref{eq:Bminiham_BW},\ref{eq:C_yangbethe_result}).

Therefore the appropriate normalization of the zero-momentum two-particle
state finite volume two-particle state is \[
|A_{1}(p)A_{1}(-p)\rangle_{L}=\frac{2}{E\sqrt{L}}\left(-\frac{1}{p}\frac{dp(L)}{dL}\right)^{1/2}|A_{1}(p)A_{1}(-p)\rangle\]
which gives\begin{eqnarray}
A(L_{0}) & = & \frac{f_{cc}}{m_{c}}\nonumber \\
B(L_{0}) & = & \sqrt{\left.\left(-\frac{1}{p}\frac{dp(L)}{dL}\right)\right|_{L=L_{0}}}\frac{2f_{c11}}{m_{c}^{3/2}}\nonumber \\
C(L_{0}) & = & \left.\left(-\frac{1}{p}\frac{dp(L)}{dL}\right)\right|_{L=L_{0}}\left(\frac{4f_{1111}}{m_{c}^{2}}+L_{0}\frac{4f_{11}}{m_{c}}\right)\label{eq:miniham_correct}\end{eqnarray}
with $p(L)$ given by (\ref{eq:quant_cond_density}). This is now
consistent with (\ref{eq:C_yangbethe_result}) and the result of the
Breit-Wigner analysis (\ref{eq:miniham_BW}). The numerical TCSA results
of section \ref{sec:Results} do indeed confirm the inclusion of the
{}``improvement'' factors related to the density of states in finite
volume. Finally, we wish to mention that similar {}``improvement''
factors have already been proposed by Lellouch and Lüscher for the
measurement of weak decay matrix element of hadrons in \cite{lellouch}.

\section{\label{sec:Numerical-methods} Numerical methods}

\subsection{\label{sub:Truncated-conformal-space} Truncated conformal space
approach}

In the framework of perturbed conformal field theory on a cylinder
with spatial circumference $L$, the action (\ref{eq:nonint_Lagrangian})
leads to the Hamiltonian\[
H=H_{*}+\mu\int_{0}^{L}dx\Phi(t,x)+\lambda\int_{0}^{L}dx\Psi(t,x)\]
where $H_{*}$ is the conformal Hamiltonian on the cylinder. Using
Euclidean time $\tau=-it$ and mapping the cylinder to the plane with
\[
z=\exp\left(\frac{2\pi(\tau-ix)}{L}\right)\quad,\quad\bar{z}=\exp\left(\frac{2\pi(\tau+ix)}{L}\right)\]
we obtain\begin{equation}
H=\frac{2\pi}{L}\left\{ \left(L_{0}+\bar{L}_{0}-\frac{c}{12}\right)+\mu\left(\frac{L}{2\pi}\right)^{1-2\Delta_{\Phi}}\int_{0}^{L}dx\Phi(z,\bar{z})+\lambda\left(\frac{L}{2\pi}\right)^{1-2\Delta_{\Psi}}\int_{0}^{L}dx\Psi(z,\bar{z})\right\} \label{eq:pcft_ham}\end{equation}
Due to translational invariance, the conformal Hilbert space $\mathcal{H}$
can be split into sectors characterized by the eigenvalues of the
total spatial momentum \[
P=\frac{2\pi}{L}\left(L_{0}-\bar{L}_{0}\right)\]
Truncating these spaces by imposing a cut in the conformal energy,
the truncated conformal space corresponding to a given truncation
reads\[
\mathcal{H}_{\mathrm{TCS}}(n,e_{\mathrm{cut}})=\left\{ |\psi\rangle\in\mathcal{H}\:|\;\left(L_{0}-\bar{L}_{0}\right)|\psi\rangle=n|\psi\rangle,\;\left(L_{0}+\bar{L}_{0}-\frac{c}{12}\right)|\psi\rangle=e|\psi\rangle\,:\, e\leq e_{\mathrm{cut}}\right\} \]
The essence of the truncated conformal space approach as introduced
by Yurov and Zamolodchikov \cite{yurov_zamolodchikov} consists in
realizing that on the space $\mathcal{H}_{\mathrm{TCS}}(n,e_{\mathrm{cut}})$
the Hamiltonian (\ref{eq:pcft_ham}) becomes a finite matrix, which
can be diagonalized numerically to get the finite volume spectrum
of the model.

In the case of the Ising model (\ref{eq:ising_action}), the Hilbert
space can be written as\[
\mathcal{H}=\bigoplus_{h=0,\frac{1}{16},\frac{1}{2}}\mathcal{V}_{h}\otimes\bar{\mathcal{V}}_{h}\]
where $\mathcal{V}_{h}$ is the irreducible Virasoro representation
of highest weight $h$ made from the Verma module\begin{equation}
V_{h}=\left\{ L_{-n_{1}}\dots L_{-n_{k}}|h\rangle:\quad L_{0}|h\rangle=h|h\rangle\right\} \label{eq:verma}\end{equation}
by factoring out the singular vectors. Using conformal Ward identities%
\footnote{The algorithm we used here was developed originally for the numerical
work in \cite{ktw}, but an explicit description of (the supersymmetric
extension of) the matrix element determination can be found in \cite{kormos}.
The only published full TCSA algorithm \cite{lassig_mussardo} is
not suitable for the present computation because it can only be applied
for low truncation levels, and is not automated enough to be scalable.%
}, the following matrices can be computed \begin{eqnarray*}
(H_{0})_{ij} & = & \left(\Delta_{i}+\bar{\Delta}_{i}-\frac{c}{12}\right)\delta_{ij}\\
G_{ij} & = & \langle i|j\rangle\\
(B_{\sigma})_{ij} & = & \langle i|\sigma(1,1)|j\rangle\\
(B_{\epsilon})_{ij} & = & \langle i|\epsilon(1,1)|j\rangle\end{eqnarray*}
where $|i\rangle$ denotes a basis of $\mathcal{H}_{\mathrm{TCS}}(n,e_{\mathrm{cut}})$.
It is very convenient to use a basis which is directly related to
the Verma module basis (\ref{eq:verma}) which is, however, not orthonormal.
Instead of performing a Gram-Schmidt orthogonalization procedure this
problem can be remedied by considering the Hamiltonian\[
(H_{TCSA})_{ij}=\frac{2\pi}{L}\left\{ \left(H_{0}\right)_{ij}+h\frac{L^{15/8}}{(2\pi)^{7/8}}(G^{-1}B_{\sigma})_{ij}+\tau L(G^{-1}B_{\epsilon})_{ij}\right\} \]
which is isospectral to (\ref{eq:pcft_ham}). We can introduce a dimensionless
Hamiltonian measuring energy and length in units given by the lowest
particle mass $m_{1}$ at the integrable point $\tau=0$\begin{eqnarray*}
h_{TCSA} & = & \frac{2\pi}{l}\left\{ H_{0}+\kappa_{h}\frac{l^{15/8}}{(2\pi)^{7/8}}\: G^{-1}B_{\sigma}+t\kappa_{h}^{8/15}l\: G^{-1}B_{\epsilon}\right\} \\
 &  & l=m_{1}L\end{eqnarray*}
where we use the parameters and notations introduced in section \ref{sub:ffpt_ising}.

The application of TCSA to perturbations of $c=1$ free boson CFT
was developed originally in \cite{SGTCSA}, and its use in double
sine-Gordon theory is described in detail in \cite{dsg,dsg_mass},
therefore we give only a brief sketch to fix our conventions. For
the case of the double sine-Gordon model (\ref{eq:dsg_action}), the
relevant Hilbert space%
\footnote{We consider only states with zero topological number, since the breather
states we are interested in can be found in this sector.%
} is\begin{equation}
\mathcal{H}=\bigoplus_{n\in\mathbb{Z}}\mathcal{F}_{n}\label{eq:sg_hilbert_space}\end{equation}
where \[
\mathcal{F}_{n}=\left\{ a_{-k_{1}}\dots a_{-k_{n}}\bar{a}_{-l_{1}}\dots\bar{a}_{-l_{m}}|n\rangle:\quad|n\rangle=\exp\left(n\frac{\beta}{2}\varphi(0,0)\right)|0\rangle\right\} \]
where the modes of the field are defined by the expansion\[
\varphi(z,\bar{z})=\varphi_{0}-\frac{i}{\sqrt{4\pi}}\Pi\log z\bar{z}+\frac{i}{\sqrt{4\pi}}\sum_{n\neq0}\frac{1}{n}\left(a_{n}z^{-n}+\bar{a}_{n}\bar{z}^{-n}\right)\]
The matrices \begin{eqnarray*}
(H_{0})_{ij} & = & \left(\Delta_{i}+\bar{\Delta}_{i}-\frac{c}{12}\right)\delta_{ij}\\
(V_{n})_{ij} & = & \frac{\langle i|\exp\left(n\frac{\beta}{2}\varphi(1,1)\right)|j\rangle}{\sqrt{\langle i|i\rangle\langle j|j\rangle}}\end{eqnarray*}
can be calculated explicitely in a closed form using the algebra of
the free field modes. The dimensionless Hamiltonian then reads\begin{eqnarray*}
h_{TCSA} & = & \frac{2\pi}{l}\left\{ H_{0}+\kappa_{\mathrm{sG}}\left(\xi\right)\frac{l^{\frac{2}{1+\xi}}}{(2\pi)^{\frac{1-\xi}{1+\xi}}}\:\frac{1}{2}\left(V_{2}+V_{-2}\right)+t\frac{l^{\frac{4+3\xi}{2+2\xi}}}{(2\pi)^{\frac{2+\xi}{2+2\xi}}}\:\frac{1}{2}\left(\mathrm{e}^{i\delta}V_{1}+\mathrm{e}^{-i\delta}V_{-1}\right)\right\} \\
 &  & l=ML\end{eqnarray*}
where $M$ is the soliton mass at the integrable point $\lambda=0$.

In both models we only consider the sector with zero total momentum
$n=0$ which is enough to obtain the necessary numerical data. The
truncation levels in both cases were such that the Hilbert space dimension
ranged from a few hundred up to 3-4000 states. This limitation was
put mainly by machine time used for the numerical diagonalization,
but in both cases the programs are fully scalable to any truncation
levels, given enough computing resources (in terms of memory and execution
time). For the Ising models we used $e_{\mathrm{cut}}=19,\dots,27$,
while in the double sine-Gordon model the typical range was $e_{\mathrm{cut}}=9,\dots,15$
for lower values of $\xi$ and $e_{\mathrm{cut}}=11,\dots,17$ for
higher values (in one case we used $e_{\mathrm{cut}}=18$). In the
case of double sine-Gordon theory, due to the choice $\delta=-\pi/2$
it is possible to project the space of states onto even and odd sectors
under the $\mathbb{Z}_{2}$ symmetry of the action (\ref{eq:dsg_action})
given by \[
\varphi\rightarrow\frac{2\pi}{\beta}-\varphi\]
as described in \cite{dsg_mass}. This reduces the dimension of the
Hilbert space with a factor of approximately $2$ which makes it possible
to use higher values for the truncation level. We used data from the
even sector to extract the results as the choice of sector does not
make much difference \cite{dsg_mass}.

\subsection{\label{sub:Testing-the-numerics} Testing the numerics and the validity
of form factor perturbation theory}

We tested the numerics by making several comparisons. First we took
the integrable point, where we checked whether we get suitable agreement
with the exact predictions of factorized scattering theory (masses,
vacuum energy density and two-particle phase shifts). 

Next, in the case of the Ising model we checked the form factor perturbation
theory predictions for the mass shifts (\ref{mass_correction}) and
the $S$ matrix correction (\ref{smatr_corr}) using the form factors
calculated by Delfino et al. in \cite{resonances} (available at \cite{http})
and formulae (\ref{mass_correction},\ref{smatr_corr}). This allowed
us to determine the range of the coupling $t$ in which perturbation
theory holds with sufficient accuracy so that we can expect reasonable
agreement for the perturbatively determined decay widths (\ref{eq:isingdecaywidth})
as well. The $S$-matrix comparison must be done carefully since formula
(\ref{smatr_corr}) is valid at a constant center of mass energy.
Therefore every formula must be converted from rapidity variables
into the energy variable as follows. The phase shift can be extracted
using\[
\delta_{\mathrm{measured}}(E)=-L\sqrt{\left(\frac{E}{2}\right)^{2}-m_{1}(\lambda)^{2}}+2n\pi\]
where $m_{1}\left(\lambda\right)$ is the mass of $A_{1}$ including
the correction to first order in $\lambda$. This must be compared
to\begin{equation}
\delta_{\mathrm{theoretical}}(E)=\delta_{0}(E)+\delta_{1}(E)\label{eq:theoretical_delta}\end{equation}
where the unperturbed phase-shift is \[
\delta_{0}(E)=-i\log S_{11}\left(\vartheta=2\,\mathrm{acosh}\frac{E}{2m_{1}}\right)\]
with $m_{1}$ denoting the mass at $\lambda=0$ since this gives the
energy dependence of the unperturbed phase-shift, and to use (\ref{smatr_corr})
it is the energy that must be kept constant. The correction term can
be written\[
\delta_{1}(E)=-i\frac{\delta S_{ab}^{cd}\left(\vartheta,\lambda\right)}{S_{11}\left(\vartheta\right)}\]
where we can use again simply\[
\vartheta=2\,\mathrm{acosh}\frac{E}{2m_{1}}\]
to first order in $\lambda$. 

\begin{figure}
\begin{center}\includegraphics[%
  scale=1.3]{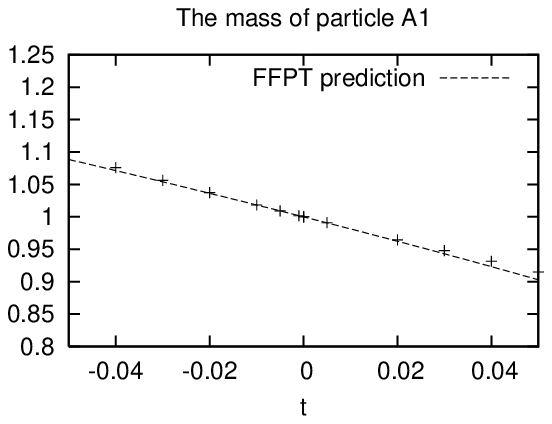}\includegraphics[%
  scale=1.3]{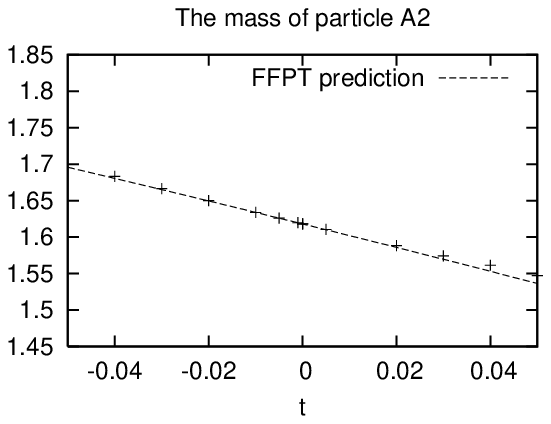}\end{center}

\caption{\label{cap:masscorcomp} Corrections to the masses of $A_{1}$ and
$A_{2}$. TCSA data are indicated with crosses (the numerical uncertainties
are too small to be displayed), while the lines give the FFPT predictions.}
\end{figure}

\begin{figure}
\begin{center}\includegraphics[%
  scale=0.6,
  angle=270]{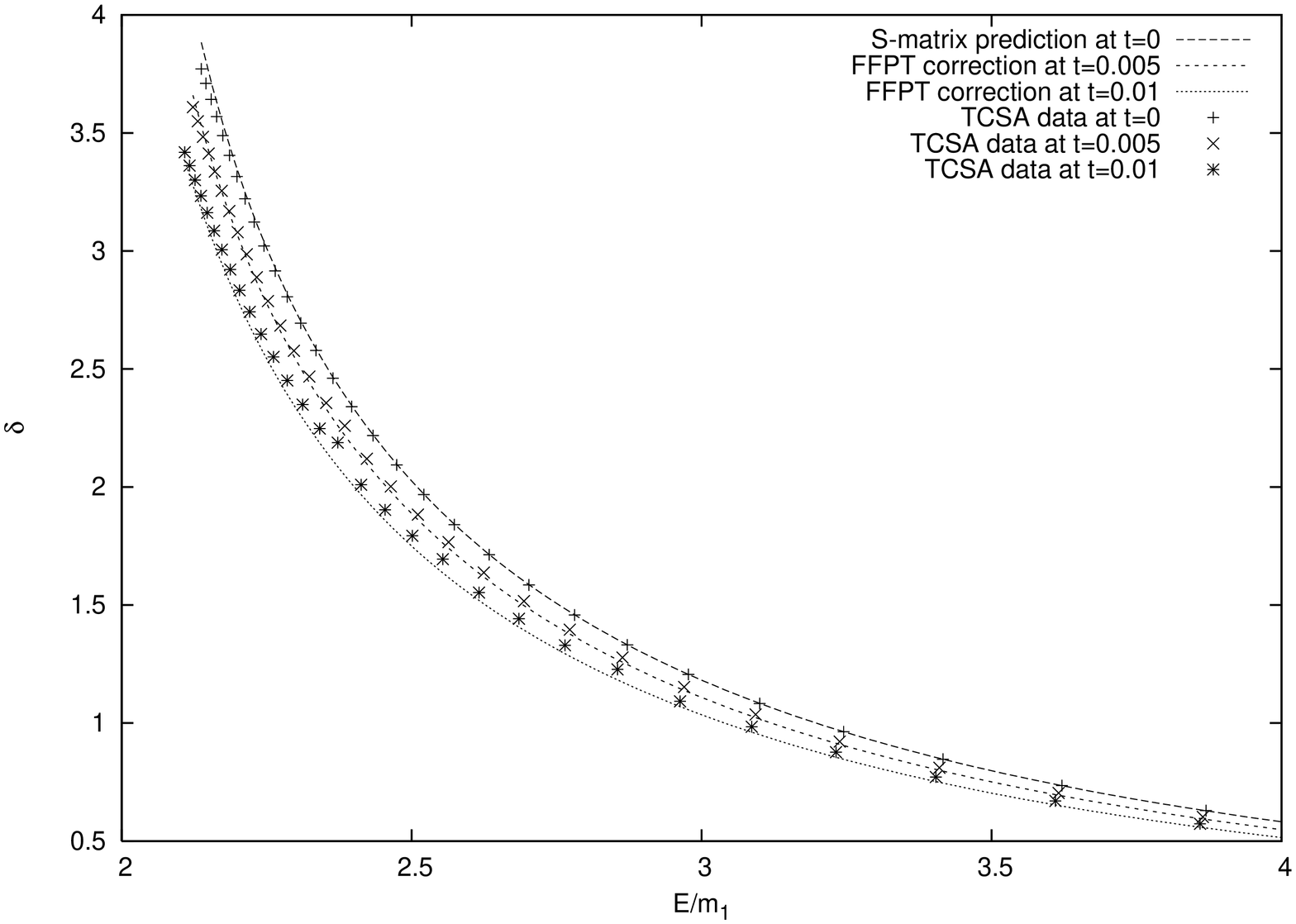}\end{center}

\caption{\label{cap:scorrcomp} The $A_{1}A_{1}$ phase shift measured at
$t=0,\,0.005,\,0.01$ and compared to the prediction (\ref{eq:theoretical_delta})
as a function of the center of mass energy $E$ (measured in units
of the unperturbed particle mass $m_{1}$). The effect of the resonance
appears on the plot as a deviation from the first order FFPT prediction
(\ref{smatr_corr}) in the range around $E_{c}\sim m_{c}\approx2.405m_{1}$,
where a {}``shoulder'' appears which can be observed quite clearly
in the $t=0.01$ data. It is a fine resolution scan of this {}``shoulder''
which is used to extract the decay width when applying the Breit-Wigner
method.}
\end{figure}

The agreement between the mass corrections and FFPT predictions for
particles $A_{1}$ and $A_{2}$ is illustrated in figure \ref{cap:masscorcomp},
while the $S$-matrix test is shown in figure \ref{cap:scorrcomp}.
Both these tests indicated that $|t|\leq0.005$ is a suitable choice
for the range of the coupling.

For the double sine-Gordon model at $\delta=-\pi/2$ it is not possible
to perform a similar test, because the mass and $S$ matrix corrections
vanish at first order in $t$ and FFPT is not formulated to second
order yet. However, the program was already thoroughly tested in previous
numerical studies \cite{dsg,dsg_mass}. We simply chose a range of
the coupling in which the measured quantities were sufficiently close
to linear dependence on $t$.

\subsection{\label{sub:Numerical-methods-decay-width} Numerical methods for
evaluation of the decay width}

We used three methods to extract the decay widths:

\begin{enumerate}
\item The Breit-Wigner analysis (see section \ref{sub:Breit-Wigner-method})
gives the decay rate as \[
\min\delta_{1}(E)-\max\delta_{2}(E)=4\sqrt{-\beta\Gamma}\]
where $\delta_{1}$ and $\delta_{2}$ are the phase shifts extracted
from the two levels using (\ref{eq:BW_phaseshift12def}) in the vicinity
of the crossing. Because we aim to extract $\sqrt{\Gamma}$ to first
order in the nonintegrable coupling $t$, $m_{a}$ and $m_{b}$ in
(\ref{eq:BW_phaseshift12def}) must be substituted with the particle
mass $m_{1}$ to first order in $t$ to keep consistency to this order.
Furthermore \[
\beta=\left.\frac{d\delta_{0}}{dE}\right|_{E=E_{0}}<0\]
can be calculated using the $t=0$ exact $S$-matrix for the background
phase shift and $E_{0}=m_{c}$ the $t=0$ particle mass since any
$t\neq0$ corrections can be consistently neglected when comparing
to the lowest order prediction (\ref{eq:decay_width}) for the decay
width $\Gamma$, from which we can extract the form factor value $f_{c11}$.
\item The {}``mini-Hamiltonian'' method (section \ref{sub:miniham_method})
when the matrix element $B$ is extracted from the minimum energy
split (\ref{eq:miniham_readout}) between the two levels \[
\delta E(L_{\mathrm{min}})=2|B\lambda|\]
In a {}``naive'' application of this method, we can then use the
naive {}``mini-Hamiltonian'' relations (\ref{eq:miniham_naive})
(or, equally well, the DGM relations (\ref{eq:miniham_DGM})) to link
this directly to the three-particle form factor:\[
B_{\mathrm{DGM}}(L_{\mathrm{min}})=\frac{2f_{c11}}{\sqrt{m_{c}^{3}L_{\mathrm{min}}}}\]

\item The improved {}``mini-Hamiltonian'' method using the relations (\ref{eq:miniham_correct})
derived in section \ref{sub:miniham_theory}\[
B(L_{0})=\sqrt{\left.-\frac{1}{p}\frac{dp(L)}{dL}\right|_{L=L_{\mathrm{min}}}}\frac{2f_{c11}}{m_{c}^{3/2}}\]
Here we replaced $L_{0}$ by $L_{\mathrm{min}}$ which is valid to
leading order in $t$. In addition, the state density correction factor
can be computed from \[
pL+\delta_{t=0}(E(p))=2n\pi\]
where $\delta_{t=0}$ is the phase shift at the integrable point $t=0$,
again to leading order in $t$.
\end{enumerate}
In each of these approaches, the numerical setting is very convenient
since the task is to find extrema of definitely convex or concave
functions (the level splitting as a function of the volume or the
phase shift as a function of energy). Therefore it is not even necessary
to find the level crossing volume $L_{0}$ very precisely: any simple
algorithm converges very fast even if the initial range given to start
the search for extremum is quite wide. The background phase shift
is given by (\ref{eq:s11_ising}) for the Ising and by (\ref{eq:s11_sg})
for the double sine-Gordon model. Both give phase shifts monotonically
decreasing with energy (in fact any two-particle scattering phases
built as products of blocks $\{ x\}$ with $0\leq x\leq1$ automatically
possess this property).

It is expected that the Breit-Wigner method should match the improved
mini-Hamiltonian method as long as the resonance is narrow enough
for the arguments of section \ref{sub:miniham_BW_link} to apply.

\subsection{\label{sub:Systematic-errors} Systematic errors}

Contrary to the situation e.g. in lattice field theory, there are
no significant sources of statistical errors. The TCSA Hamiltonian
matrices are diagonalized numerically with double (16 digits) precision,
and any error in calculating their spectra can be considered minuscule.
There are, however, systematic errors in the extracted decay widths
because (1) the TCSA Hamiltonian is not the exact one of the finite
size system and (2) the analysis in Section \ref{sec:resonance_finvol}
neglected residual finite size corrections (i.e. those that decay
exponentially with the volume).

The first source goes by the name {}``truncation errors'', which
originate from the fact that TCSA neglects an infinite tower of states
lying above the truncation level, and grow with the volume. In the
region close to a line crossing they manifest themselves most prominently
in the fact that even for $t=0$ the degeneracy is generally not exact.
This can be modeled in the {}``mini-Hamiltonian'' picture by adding
a correction matrix representing the truncation errors to (\ref{eq:miniham_readout}\ref{eq:miniham})\begin{equation}
H=E_{0}+\left(L-L_{0}\right)\left(\begin{array}{cc}
\alpha_{1}\\
 & \alpha_{2}\end{array}\right)+\lambda\left(\begin{array}{cc}
A(L) & B(L)\\
B(L) & C(L)\end{array}\right)+\left(\begin{array}{cc}
\delta E_{0}+a & b\\
b & \delta E_{0}-a\end{array}\right)\label{eq:miniham_trunc}\end{equation}
As a result, the minimal splitting changes to \begin{equation}
\delta E(L_{\mathrm{min}})=2|b+B\lambda|=2\left|B(\lambda-\lambda_{0})\right|\label{eq:residual_splitting}\end{equation}
but $B$ can still be read out from the slope of the dependence of
$\delta E(L_{\mathrm{min}})$ on $\lambda$ and the form factor amplitude
$f_{c11}$ can then be determined from \[
B=\frac{2f_{c11}}{\sqrt{m_{c}^{3}L_{0}}}\]
or\[
B=\sqrt{\left.-\frac{1}{p}\frac{dp(L)}{dL}\right|_{L=L_{\mathrm{0}}}}\frac{2f_{c11}}{m_{c}^{3/2}}\]
depending on whether we use the naive or improved {}``mini-Hamiltonian''
method, respectively.

\begin{figure}
\begin{center}\includegraphics[%
  scale=0.5,
  angle=270]{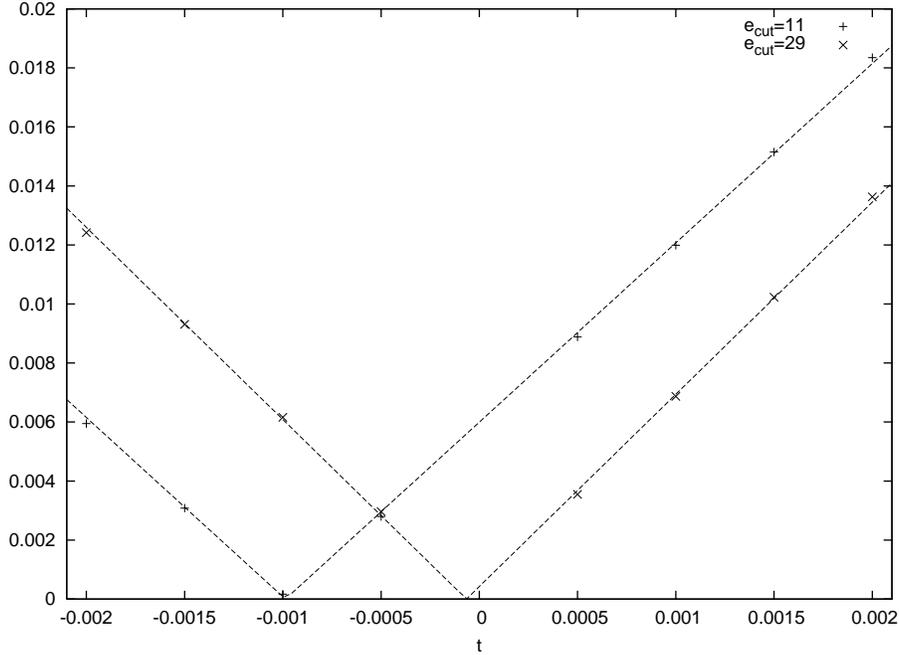}\end{center}

\caption{\label{cap:tfit} The Breit-Wigner phase shift splitting $\min\delta_{1}(E)-\max\delta_{2}(E)$
as a function of the dimensionless coupling $t$ in the Ising case
for $e_{\mathrm{cut}}$ values of $11$ and $29$. }
\end{figure}

This effect is demonstrated in figure \ref{cap:tfit}, where it is
obvious that the residual splitting decreases with increasing truncation
level and gives an idea of how precise the effective description (\ref{eq:miniham_trunc})
really is. The effect of the other parameter $a$ is to shift the
value of $L_{\mathrm{min}}$ while $\delta E_{0}$ is the truncation
error in determining the resonant energy $E_{0}$.

It is easy to see that the same modification must be performed when
applying the Breit-Wigner method, and relation (\ref{eq:BW_readout})
must be altered as \[
\min\delta_{1}(E)-\max\delta_{2}(E)=4\left|B'(\lambda-\lambda_{0})\right|\]
from which the form factor amplitude $f_{c11}$ for the process $A_{c}\rightarrow A_{a}+A_{a}$
can be read off using\[
B'=\sqrt{-\left.\frac{d\delta_{0}}{dE}\right|_{E=E_{0}}}\frac{f_{c11}}{m_{c}m_{1}^{1/2}\left(\left(\frac{m_{3}}{2m_{1}}\right)^{2}-1\right)^{1/4}}\]

In the sine-Gordon case, numerics always yields $\lambda_{0}=0$ i.e.
there is no residual splitting between the numerically determined
levels at the integrable point. The reason is that the sine-Gordon
spectrum is unaffected by $\lambda\rightarrow-\lambda$ since this
can be implemented by shifting $\varphi\rightarrow\varphi+2\pi/\beta$
in the action (\ref{eq:dsg_action}) and this symmetry is respected
by the truncation procedure, so the numerical spectrum is even in
$\lambda$ (up to very small errors in numerical matrix diagonalization).

For a fixed value of $e_{\mathrm{cut}}$ and a choice of level crossing,
we can then determine a value for $f_{c11}$. Then the next task is
to extrapolate the results to infinite truncation level and then eliminate
finite size effects (extrapolate to infinite volume). 

Since truncation errors grow with increasing volume while finite size
errors become larger with decreasing volume, some compromise must
be struck in between. Masses are normally measured by finding the
region where the measured gap is closest to being constant (has a
minimum slope) and taking the value of the gap there. Two-particle
phase shifts are measured over an extended volume range, and so they
are affected by truncation errors at small, and by finite size errors
at large values of the center of mass energy, and therefore it is
very hard to perform any sort of optimization (except for a choice
of the two-particle level used to extract the phase shift). Decay
widths are measured in small regions around level crossings, and we
perform the measurement on the first few of them, for several different
values of $e_{\mathrm{cut}}$.

\begin{figure}
\begin{center}\subfigure[improved "mini-Hamiltonian" method, $R=2.5$]{\includegraphics[%
  scale=0.6,
  angle=270]{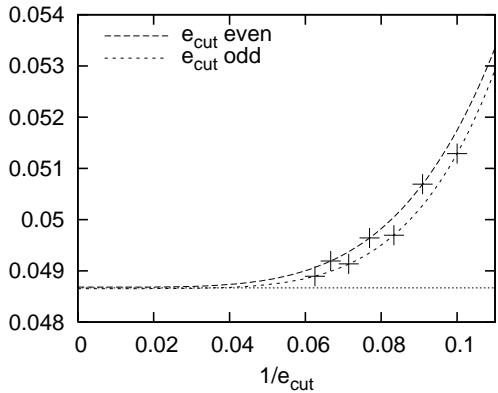}}\subfigure[Breit-Wigner method, $R=1.7$]{\includegraphics[%
  scale=0.6,
  angle=270]{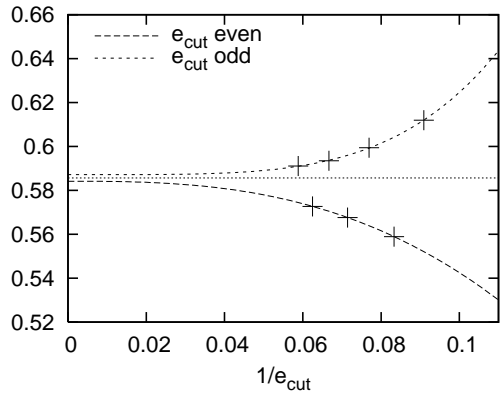}}\end{center}

\caption{\label{cap:Truncation-level-extrapolation}Truncation level extrapolation
for $s_{311}$ in the double sine-Gordon model ($R$ is related to
$\xi$ via $2R^{2}=1+\xi^{-1}$)}
\end{figure}

Truncation errors can be improved by fitting the truncation level
dependence by some function for extrapolation to $e_{\mathrm{cut}}=\infty$.
Detailed examination of numerical data shows that they can be fit
in most cases by a function of the sort\begin{equation}
f\left(e_{\mathrm{cut}},L\right)=f(L)+a(L)e_{\mathrm{cut}}^{-x}\label{eq:ecut_extrapol}\end{equation}
where $f(L)$ is the extrapolated value of the measured physical quantity
$f$ at volume $L$, and the second terms describes the truncation
effects with $a$ and the exponent $x>0$ to be determined from the
fit. Such an extrapolation was used in \cite{kfold} where it proved
very useful in the determination of vacuum expectation values. Experimenting
with this technique in cases where a theoretical prediction for the
measured quantity is available shows that this really helps, and in
many cases the truncation errors are improved by an order of magnitude.
Based on the numerical results the exponent $x$ can be supposed to
be independent of the coupling $t$ and depends only very mildly on
$L$ (in the sine-Gordon case, however, a dependence on the parameter
$\xi$ must be taken into account). For an illustration see figure
\ref{cap:Truncation-level-extrapolation} which shows that separate
extrapolation is necessary for odd and even values of the truncation
level. The reason is that only when increasing the truncation level
by a step of $2$ we get new vectors in \emph{each} (Verma or Fock)
module: since the momentum is zero, we must increase the descendent
quantum numbers on both left and right by $1$, and thus it is only
consistent to group together data pertaining to $e_{\mathrm{cut}}$
that differ in steps of $2$. In the sine-Gordon case the situation
is further complicated by the fact that the number of Fock modules
$\mathcal{F}_{n}$ contributing states in (\ref{eq:sg_hilbert_space})
also grows with the truncation level, but this effect turns out to
be numerically irrelevant. 

Unfortunately, there is no theoretical method providing a reliable
estimate of truncation errors. However, in the case of the double
sine-Gordon model the two Breit-Wigner extrapolation curves approach
the average value from the two sides (as in figure \ref{cap:Truncation-level-extrapolation}
(b)) and therefore residual truncation errors can be estimated by
the difference between the two extrapolated values. We adopt the same
procedure for the {}``mini-Hamiltonian'' methods as well, although
in that case the situation is less clear-cut because the two sets
of data approach the extrapolated limit from the same side. In the
case of the Ising model, the even/odd extrapolations approach the
final value from the same side for both methods, so the difference
between is less reliable as an estimate for the truncation errors.

The other source of systematic errors is \emph{residual} finite size
corrections originating from vacuum polarization and the finite range
of interactions. According to Lüscher's result \cite{luscher_2particle,luscher_1particle}
these are suppressed by a factor of the form $\exp(-ML)$ where $M$
is some characteristic mass scale (typically the mass of the lightest
particle). Such effects are e.g. the volume dependence of the particle
masses and vacuum (Casimir) energy. These were neglected in all the
considerations in Section \ref{sec:resonance_finvol}, where the only
finite size effects taken into account are the ones that decay as
a power (generally $L^{-2}$) of the volume. 

Having taken care of truncation errors, residual finite size corrections
can in principle be suppressed by fitting an appropriate extrapolation
function. Suppose that the decay width can be calculated from several
crossings which take place in very small regions around different
values of the volume $L_{1},\dots,L_{n}$ and that truncation errors
in these determinations can be kept suitably small. Using our general
understanding of finite size effects, we can think of the decay form
factor $f_{c11}$ determined at a crossing at $L_{i}$ as the value
of a function $f_{c11}(L)$ at $L=L_{i}$ with the property that\[
f_{c11}(L)=f_{c11}+O\left(\mathrm{e}^{-ML}\right)\]
where $f_{c11}$ is the value predicted by the infinite volume quantum
field theory. Unfortunately our numerical results (even after extrapolation
in the truncation level) do not have sufficient precision to fit the
exponential term in a reliable way. Therefore we will not perform
this extrapolation in the sequel, and restrict ourselves to quoting
the values of $f_{c11}$ as measured at each level crossing separately
in the Ising case; in the sine-Gordon case truncation errors allowed
us to use only the first level crossing, i.e. the one that occurs
at the smallest value of the volume.

\section{\label{sec:Results} Results}

\subsection{\label{sub:results_Ising} Ising model}

\subsubsection{\label{sub:ising_C_test} Testing the mini-Hamiltonian coefficient
$C$}

It is interesting to check whether the improved {}``mini-Hamiltonian''
description (\ref{eq:miniham_correct}) is consistent with the numerical
data. The correct formula for $B$ can be tested by comparing the
results for the decay width extracted using the Breit-Wigner method,
and will be preformed in subsection \ref{sub:ising_decaywidths} below.
Here we perform a quick test for the coefficient $C$. Let us take
the sum of the two energy levels $E_{1}(L)+E_{2}\left(L\right)$ at
the volume where the level crossing occurs at $\lambda=0$. Then using
the results of subsection (\ref{sub:miniham_BW_link}) this must be
equal to \[
E_{1}(L_{0})+E_{2}(L_{0})=2m_{c}+\lambda(A(L_{0})+C(L_{0}))\]
when taken at $L=L_{0}$ where $L_{0}$ is the volume where the levels
cross. Using TCSA data with truncation level $e_{\mathrm{cut}}=27$,
we measured this energy at five different values of the integrability
breaking coupling $t=-0.003,-0.001,0,0.001,0.003$, at the the first
three level crossing. The results were then fitted by a linear function,
parameterized as \[
\frac{E_{11}(L_{0})+E_{c}(L_{0})}{m_{1}}=a+bt\]
and are summarized in table \ref{cap:testing_c}. The theoretical
value of $a$ is \[
a=\frac{2m_{4}}{m_{1}}=4.80973\dots\]

\begin{table}
\begin{center}\begin{tabular}{|c||c|c||c|c|c|}
\hline 
&
\multicolumn{2}{c||}{Numerics (TCSA)}&
\multicolumn{3}{c|}{{}``mini-Hamiltonian'' prediction}\tabularnewline
\hline 
$L_{0}$&
$a$&
$b$&
$b$ (DGM)&
$b$ (naive)&
$b$ (improved) \tabularnewline
\hline
\hline 
18.152&
$4.52877$&
$-3.598$&
$-0.509$&
$-3.580$&
$-3.974$\tabularnewline
\hline 
24.900&
$4.80039$&
$-4.348$&
$-0.947$&
$-4.016$&
$-4.364$\tabularnewline
\hline 
34.184&
$4.81197$&
$-4.586$&
$-1.263$&
$-4.334$&
$-4.616$\tabularnewline
\hline
\end{tabular}\end{center}

\caption{\label{cap:testing_c} Testing the effect of the disconnected term
and the finite volume state density factor on the {}``mini-Hamiltonian''
description. We compare the TCSA data to three alternatives: the DGM
result (\ref{eq:miniham_DGM}), the naive {}``mini-Hamiltonian''
(\ref{eq:miniham_naive}) and the improved version (\ref{eq:miniham_correct}).
Due to residual finite size effects, the comparison between the naive
and the improved {}``mini-Hamiltonian'' at the first level crossing
cannot be taken seriously.}
\end{table}

It is obvious that to get agreement with the TCSA results, both the
disconnected term and the correct finite volume normalization of two-particle
states must be taken into account. Another important point is that
residual finite size effects are largest at the first level crossing,
where $a=4.52877$ is quite far from the theoretical value, therefore
one cannot distinguish between the naive and the improved {}``mini-Hamiltonian''
methods. On the other hand, the results at the other two level crossings
show that both the disconnected terms and the improvement is needed
in order to get appropriate agreement with the numerical data.

\subsubsection{\label{sub:ising_decaywidths} Decay widths}

We measured decay width for the process $A_{4}\rightarrow A_{1}+A_{1}$
at three level crossings, while $A_{5}\rightarrow A_{1}+A_{1}$ was
measured at one level crossing. We extracted the form factor matrix
elements $f_{411}$ and $f_{511}$ using all the three methods ({}``naive''
mini-Hamiltonian -- equivalent with the DGM one for the case of the
decay width--, improved mini-Hamiltonian and Breit-Wigner method)
and the end results (after truncation level extrapolation) are presented
in tables \ref{cap:411} and \ref{cap:511}. 

From table \ref{cap:411} we can see that indeed it is the improved
{}``mini-Hamiltonian'' method (rather than the naive one) which
is consistent with the Breit-Wigner approach. It is also apparent
that the extracted values do change with the volume, and that the
naive {}``mini-Hamiltonian'' method significantly overshoots the
theoretical prediction, while both the improved {}``mini-Hamiltonian''
and Breit-Wigner methods are in reasonable agreement with it (within
1\% at the 3rd level crossing).

\begin{table}
\begin{center}\begin{tabular}{|c|c|c|c|c|}
\hline 
$n$&
$m_{1}L_{0}$&
naive mH (o/e)&
improved mH (o/e)&
Breit-Wigner (o/e)\tabularnewline
\hline
\hline 
1&
18.152&
37.658/37.662&
33.422/33.426&
33.255/33.336\tabularnewline
\hline 
2&
24.900&
38.871/38.939&
35.736/35.799&
34.574/34.887\tabularnewline
\hline 
3&
34.184&
39.099/{*}&
36.829/{*}&
36.318/{*}\tabularnewline
\hline
\end{tabular}\end{center}

\caption{\label{cap:411} The measured values of $f_{411}$ where $n$ is
the label of the level crossing, $L_{0}$ is the place of the level
crossing for $t=0$ and the last three columns give the result extracted
using the method indicated at the top. The two numbers given are the
extrapolated values in the truncation level for $e_{\mathrm{cut}}=$odd/even
($*$s indicate the cases where the data could not be reliably extrapolated).
The theoretical prediction is $f_{411}=36.730$. }
\end{table}

In the case of the matrix element $f_{511}$ truncation errors are
expected to be higher because the levels lie much higher in the spectrum.
This observation is borne out by the data presented in table \ref{cap:511},
where we gave the data at the two largest values of $e_{\mathrm{cut}}$
instead of extrapolating them in the truncation level, because in
most cases the extrapolation function (\ref{eq:ecut_extrapol}) could
not be fitted in a reliable way. The variation of the primary quantities
$\Delta\delta$ used in the Breit-Wigner method and the minimum energy
split used in both mini-Hamiltonian methods as a function of $t$
also shows much larger $t^{2}$ effects (but decreasing $t$ is not
possible beyond a certain limit because of the presence of the residual
line splitting \ref{eq:residual_splitting}). Therefore the matching
between the Breit-Wigner and the improved {}``mini-Hamilton'' methods
is less precise, but the latter is still somewhat closer to the Breit-Wigner
result than the naive version. Despite the much higher errors the
two best estimates $20.349$ (from the improved {}``mini-Hamiltonian''
method) and $18.612$ (from the Breit-Wigner method) are still within
a few percent of the theoretical value $19.163$.

\begin{table}
\begin{center}\begin{tabular}{|c|c|c|c|c|}
\hline 
$n$&
$m_{1}L_{0}$&
naive mH (27/28)&
improved mH (27/28)&
Breit-Wigner (27/28)\tabularnewline
\hline
\hline 
1&
23.206&
21.011/21.120&
20.244/20.349&
18.551/18.612\tabularnewline
\hline
\end{tabular}\end{center}

\caption{\label{cap:511} The measured values of $f_{511}$ at the first level
crossing. $L_{0}$ is the place of the level crossing for $t=0$ and
the last three columns give the result extracted using the method
indicated at the top. The theoretical prediction is $f_{511}=19.163$.
Data at the second and higher crossings already have too large truncation
errors. Some of the data could not be extrapolated in the truncation
level; therefore we only quote the values measured at $e_{\mathrm{cut}}=27$
and $28$. }
\end{table}

\subsection{\label{sub:results_dsg} Double sine-Gordon model}

For convenience, we parameterize the sine-Gordon coupling as follows\[
\beta=\frac{\sqrt{4\pi}}{R}\quad,\quad\xi=\frac{1}{2R^{2}-1}\]

\begin{table}
\begin{center}\begin{tabular}{|c|c|c|c|c|c|c|}
\hline 
$R$&
nmH&
imH&
BW (e/o)&
FFPT&
$ML_{0}$&
$\mathrm{e}^{-m_{1}L_{0}}$\tabularnewline
\hline
\hline 
1.44&
$0.931\pm0.025$&
$0.821\pm0.021$&
{*}/{*}&
0.7759&
16.469&
$10^{-7}$\tabularnewline
\hline 
1.5&
$1.328\pm0.038$&
$1.185\pm0.034$&
{*}/{*}&
1.1694&
12.398&
$2\cdot10^{-6}$\tabularnewline
\hline 
1.6&
$0.959\pm0.012$&
$0.883\pm0.011$&
0.904/0.893&
0.9303&
11.588&
0.0002\tabularnewline
\hline 
1.7&
$0.630\pm0.004$&
$0.592\pm0.003$&
0.584/0.587&
0.6383&
11.927&
0.0005\tabularnewline
\hline 
1.9&
$0.286\pm0.001$&
$0.274\pm0.001$&
{*}/0.269&
0.2917&
13.615&
0.0011\tabularnewline
\hline 
2.2&
$0.1076\pm0.0002$&
$0.1046\pm0.0002$&
0.110031/0.110031&
0.0999&
17.475&
0.0018\tabularnewline
\hline 
2.5&
$0.04867\pm2\cdot10^{-5}$&
$0.04767\pm2\cdot10^{-5}$&
{*}/0.049951&
0.0392&
22.309&
0.0023\tabularnewline
\hline 
2.6&
$0.03845\pm6\cdot10^{-5}$&
$0.03773\pm6\cdot10^{-5}$&
0.041271/0.041283&
0.0295&
24.089&
0.0024\tabularnewline
\hline 
2.7&
$0.03075\pm5\cdot10^{-5}$&
$0.03022\pm5\cdot10^{-5}$&
0.033467/0.033403&
0.0224&
25.946&
0.0025\tabularnewline
\hline
\end{tabular}\end{center}

\caption{\label{cap:s311_measured} Measured values of $s_{311}$ using the
{}``mini-Hamiltonian'' and Breit-Wigner methods. The column label
{}``nmH'' corresponds to the naive, while {}``imH'' to the improved
{}``mini-Hamiltonian'' method. The next column contain the results
of the Breit-Wigner method extrapolated for even/odd truncation levels,
respectively, which also give a relatively good estimate of residual
truncation errors. FFPT labels the theoretical prediction, and the
last two columns contain the volume corresponding to the level crossing
at $t=0$ and the characteristic suppression factor of the leading
finite size correction. The $*$'s label the cases where the Breit-Wigner
data could not be extrapolated meaningfully in the truncation level.}
\end{table}

\begin{figure}
\begin{center}\includegraphics{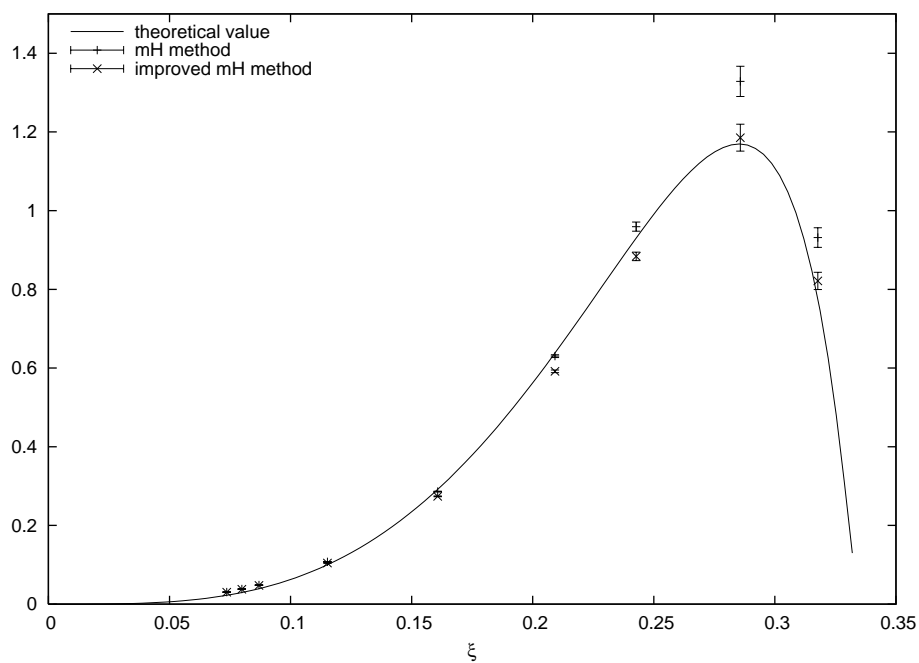}\end{center}

\caption{\label{cap:s311_plot} The theoretical value of $s_{311}$ plotted
against $\xi$, together with the numerical results obtained with
the naive and the improved {}``mini-Hamiltonian'' methods (Breit-Wigner
data are omitted, since they are not significantly different from
the latter). The error bars include only the residual truncation errors
(the residual finite size errors are not estimated).}
\end{figure}

In table \ref{cap:s311_measured} we compare the measured values of
$s_{311}$ (see eq. (\ref{eq:sg_decaywidth})) for the three methods
with the predictions, using the first level crossing (for a plot,
see figure \ref{cap:s311_plot}). For the {}``mini-Hamiltonian''
methods, the values in the table are the averages of the two separate
extrapolations for $e_{\mathrm{cut}}=$even/odd, while the errors
shown are the estimated residual truncation errors calculated dividing
the difference of the two extrapolated values by $2$. Because the
truncation level extrapolation for the Breit-Wigner method is more
problematic (at least for some values of $R$), we show the results
for even/odd truncation levels separately. Truncation errors increase
with decreasing $R$, and indeed this is most prominently shown by
the relative deviation of the measured and theoretical value at $R=1.44$
where the Breit-Wigner method is not applicable due to the fact that
albeit theoretically the condition of the monotonous decrease of the
background phase shift\[
\frac{d\delta_{0}}{dE}<0\]
 is still satisfied, the numerical data fail to have this property.
In addition, the value of $L_{0}$ for $R=1.44$ is also significantly
larger than for $R=1.5$, which increases truncation errors too. For
larger $R$ the deviation between TCSA numerics and FFPT seems to
be controlled by the residual finite size effects. Higher level crossing
show too large truncation effects, and are not useful for numerical
analysis. Therefore there are not enough data to estimate the residual
finite size effects, and to get some idea of their magnitude we quote
the exponential suppression factor characterizing their decay with
the volume. Note that this is not a very precise estimate since there
must certainly be additional dependence on $\xi$ and on some power
of $L_{0}$, the form of which is not known. However, it captures
the tendency correctly: the residual finite size corrections increase
with $R$, again making the measurements at higher values of $R$
less precise. 

It is also obvious that the improvement of the {}``mini-Hamiltonian''
method is most important for smaller $R$ and for $R=1.5\dots1.9$
it is required to get appropriate agreement with the Breit-Wigner
method. For larger values of $R$ the improvement is too small to
draw any useful distinction between the naive and the improved methods.
At $R=1.44$ we see that despite the problems with the Breit-Wigner
method, the improved {}``mini-Hamiltonian'' method is still applicable
and is in reasonable agreement with the theoretical prediction (relative
deviation 6\%), while the naive (DGM) version is again off the mark
with a relative deviation of 20\%. 

On the other hand, note that the deviations in the range $R=1.44\dots2.2$
obtained at the \emph{first} level crossing are comparable in magnitude
to those obtained in the case of the Ising model for the process $A_{4}\rightarrow A_{1}+A_{1}$
at the \emph{second} level crossing and so we consider the agreement
with the theoretical predictions fully satisfactory.

\section{\label{sec:Conclusions} Conclusions and outlook}

In this paper we investigated the signatures of resonances in finite
volume quantum field theory. The central result of the paper is the
development of two methods: (a) the Breit-Wigner method utilizing
the standard parameterization of the resonant contribution to the
two-particle phase shift, and (b) the improved {}``mini-Hamiltonian''
method based on the explicit Hamiltonian description of quantum field
theory, to extract (partial) decay widths of particles given the finite
volume spectrum. 

We have shown the equivalence between these methods to first order
in the resonance width (i.e. for narrow resonances). As a further
result of this investigation, we now have a finite size description
of the resolution of level crossings which brings Lüscher's description
of finite size effects, the Breit-Wigner parameterization of resonances,
the {}``mini-Hamiltonian'' method and form factor perturbation theory
together in a consistent framework.

Both methods are extremely simple to implement numerically since all
what is needed is to find a minimum/maximum of some convex/concave
function (the level splitting as a function of the volume or the phase
shifts extracted from the two levels as a function of energy). However,
the Breit-Wigner method does not seem to be as robust numerically
under extrapolation in the truncation level (i.e. the UV cut-off)
as the improved {}``mini-Hamiltonian'' method, so the latter appears
to be preferable. 

Using the machinery available in the realm of 1+1 dimensional quantum
field theories, namely form factor perturbation theory and truncated
conformal space approach we demonstrated the efficiency of these methods,
by showing agreement between the numerically extracted decay matrix
elements and the theoretical ones in many cases up to a precision
of 1 to 7\% (as shown in figure \ref{cap:s311_plot}). The most important
factor in achieving this level of precision this is that both methods
measure directly effects proportional to $\sqrt{\Gamma}$ (where $\Gamma$
is the resonance width) in contrast to other methods proposed in the
literature (such as the {}``plateau slope'' method in \cite{luscher_resonance})
which attempt to detect effects of order $\Gamma$. Therefore, our
methods are much more sensitive when the resonance is narrow. Another
important advantage of methods (a) and (b) is that they only need
data from a small vicinity of the level crossings resolved by the
finite width of the resonance, while (as discussed at the end of subsection
\ref{sub:Breit-Wigner-method}) fitting a Breit-Wigner shape directly
to the extracted phase shift is not as reliable because it uses data
from an extended range of the volume (containing two subsequent level
crossings), and the residual finite size effects introduce significant
distortion in the shape of the resonance curve.

It is clear that since the necessary description of finite size effects
exists in arbitrary dimensions \cite{luscher_1particle,luscher_2particle,luscher_resonance},
the development of the two methods in Section \ref{sec:resonance_finvol}
can be generalized to any number of space-time dimensions in a straightforward
way and the results of the present work can be extended to more interesting
theories formulated e.g. on the lattice. We remark that even the form
factor approach to the calculation of the decay width has its $3+1$
dimensional counterpart in the description of the weak decays of hadrons,
where the appropriate form factor is the hadronic matrix element of
the weak current. The crucial issue is whether the necessary precision
in the measurement of finite size effects can be attained. Therefore
the most important open problem is to extend these results to theories
in physical ($3+1$) space-time dimensions, and in particular to phenomenologically
relevant models.

\subsection*{Acknowledgments}

G.T. would like to thank Z. Bajnok for useful discussions and for
drawing attention to paper \cite{resonances}. This work was partially
supported by the EC network {}``EUCLID'', contract number HPRN-CT-2002-00325,
and Hungarian research funds OTKA T043582, K60040 and TS044839. G.T.
was also supported by a Bolyai J{\'a}nos research scholarship.

\appendix

\makeatletter \renewcommand{\theequation}{\hbox{\normalsize\Alph{section}.\arabic{equation}}} \@addtoreset{equation}{section} \renewcommand{\thefigure}{\hbox{\normalsize\Alph{section}.\arabic{figure}}} \@addtoreset{figure}{section} \renewcommand{\thetable}{\hbox{\normalsize\Alph{section}.\arabic{table}}} \@addtoreset{table}{section} \makeatother

\section{\label{sec:s311_appendix} The 311 form factor}

In this paper we only calculate the decay width for the simplest possible
case $B_{3}\rightarrow B_{1}+B_{1}$. In order to accomplish this,
we need the form factor of the perturbing field $\Psi=\cos\left(\frac{\beta}{2}\Phi+\delta\right)$\[
F_{311}^{\Psi}\left(\vartheta_{1},\vartheta_{2},\vartheta_{3}\right)=\langle0|\cos\left(\frac{\beta}{2}\Phi+\delta\right)|B_{3}(\vartheta_{1})B_{1}(\vartheta_{2})B_{1}(\vartheta_{3})\rangle\]
In \cite{formfactors} Lukyanov obtained a closed expression for the
form factors of exponential operators with any number of $B_{1}$-s
in the asymptotic state. It can be written as an expectation value\[
F_{1\dots1}^{a}\left(\vartheta_{n},\dots,\vartheta_{1}\right)=\langle0|\exp\left(ia\Phi\right)|B_{1}(\vartheta_{n})\dots B_{1}(\vartheta_{1})\rangle=\mathcal{G}_{a}\langle\Lambda(\vartheta_{n})\dots\Lambda(\vartheta_{1})\rangle\]
where\begin{eqnarray*}
 &  & \Lambda(\vartheta)=\frac{\bar{\lambda}\left(\xi\right)}{2\sin\pi\xi}\left(e^{-i\frac{a\pi\xi}{\beta}}e^{-i\omega(\vartheta+i\frac{\pi}{2})}-e^{i\frac{a\pi\xi}{\beta}}e^{i\omega(\vartheta-i\frac{\pi}{2})}\right)\\
 &  & \bar{\lambda}\left(\xi\right)=2\cos\left(\frac{\pi\xi}{2}\right)\sqrt{2\sin\left(\frac{\pi\xi}{2}\right)}\exp\left(-\int_{0}^{\pi\xi}\frac{dt}{2\pi}\frac{t}{\sin t}\right)\end{eqnarray*}
and the expectation value can be computed using Wick's theorem for
the generalized free field $\omega$ (in a multiplicative form) with
\begin{eqnarray*}
 &  & \langle e^{i\alpha\omega(\vartheta)}\rangle=1\\
 &  & \langle\prod_{j=1}^{N}e^{i\alpha_{j}\omega(\vartheta_{j})}\rangle=\prod_{j=1}^{N-1}\prod_{k=j+1}^{N}R\left(\vartheta_{k}-\vartheta_{j}\right)^{\alpha_{k}\alpha_{j}}\\
\end{eqnarray*}
where \begin{eqnarray}
R\left(\vartheta\right) & = & \mathcal{N}\exp\left\{ 8\int_{0}^{\infty}\frac{dt}{t}\frac{\sinh t\sinh t\xi\sinh t(1+\xi)}{\sinh^{2}2t}\sinh^{2}t\left(1-\frac{i\vartheta}{\pi}\right)\right\} \label{eq:R_intrep}\\
\mathcal{N} & = & \exp\left\{ 4\int_{0}^{\infty}\frac{dt}{t}\frac{\sinh t\sinh t\xi\sinh t(1+\xi)}{\sinh^{2}2t}\right\} \nonumber \end{eqnarray}
(the integral representation is only valid in the strip $-2\pi+\pi\xi<\Im m\vartheta<-\pi\xi$)
is the so-called minimal $B_{1}B_{1}$ form factor satisfying \begin{eqnarray}
R(-\vartheta) & = & S_{11}(\vartheta)R(\vartheta)\nonumber \\
R(-i\pi+\vartheta) & = & R(-i\pi-\vartheta)\label{eq:minff_defining_relations}\end{eqnarray}
where $S_{11}(\vartheta)$ is the \textbf{$B_{1}-B_{1}$} two-particle
scattering amplitude. Furthermore\begin{eqnarray}
\mathcal{G}_{a}(\beta)=\langle e^{ia\Phi}\rangle & = & \left[\frac{M\sqrt{\pi}\Gamma\left(\frac{4\pi}{8\pi-\beta^{2}}\right)}{2\Gamma\left(\frac{\beta^{2}/2}{8\pi-\beta^{2}}\right)}\right]^{\frac{a^{2}}{4\pi}}\label{lz_vev}\\
 & \times & \exp\left\{ \int_{0}^{\infty}\frac{dt}{t}\left[\frac{\sinh^{2}\left(\frac{a\beta}{4\pi}t\right)}{2\sinh\left(\frac{\beta^{2}}{8\pi}t\right)\cosh\left(\left(1-\frac{\beta^{2}}{8\pi}\right)t\right)\sinh t}-\frac{a^{2}}{4\pi}e^{-2t}\right]\right\} \nonumber \end{eqnarray}
is the exact vacuum expectation value of the exponential field \cite{exact_vevs}.

Performing the Wick contractions explicitly we obtain\begin{eqnarray*}
\langle\Lambda(\vartheta_{1})\dots\Lambda(\vartheta_{n})\rangle & = & \left(-\frac{\bar{\lambda}}{2\sin\pi\xi}\right)^{n}\sum_{\{\alpha_{j}=\pm1\}}\Bigg\{\left(\prod_{j=1}^{n}\alpha_{j}e^{\alpha_{j}i\frac{a\pi\xi}{\beta}}\right)\times\\
 &  & \prod_{j=1}^{n-1}\prod_{k=j}^{n}R\left(\vartheta_{k}-\vartheta_{j}-(\alpha_{k}-\alpha_{j})i\frac{\pi}{2}\right)^{\alpha_{k}\alpha_{j}}\Bigg\}\end{eqnarray*}
where the sum goes over all the $2^{n}$ possible {}``configurations''
of the auxiliary variables $\alpha_{j}=\pm1$. Using the identity\begin{equation}
R(\vartheta)R(\vartheta\pm i\pi)=\frac{\sinh\vartheta}{\sinh\vartheta\mp i\sin\pi\xi}\label{eq:R_ipi_identity}\end{equation}
we can write \[
R\left(\vartheta_{k}-\vartheta_{j}-(\alpha_{k}-\alpha_{j})i\frac{\pi}{2}\right)^{\alpha_{k}\alpha_{j}}=\left(1+i\frac{\alpha_{k}-\alpha_{j}}{2}\frac{\sin\pi\xi}{\sinh\vartheta}\right)R\left(\vartheta_{k}-\vartheta_{j}\right)\]
The end result is\begin{eqnarray*}
F_{1\dots1}^{a}\left(\vartheta_{1},\dots,\vartheta_{n}\right) & = & \mathcal{G}_{a}\left(-\frac{\bar{\lambda}}{2\sin\pi\xi}\right)^{n}P_{n}^{a}\left(\vartheta_{n},\dots,\vartheta_{1}\right)\prod_{j=1}^{n-1}\prod_{k=j+1}^{n}R\left(\vartheta_{k}-\vartheta_{j}\right)\\
P_{n}^{a}\left(\vartheta_{1},\dots,\vartheta_{n}\right) & = & \sum_{\{\alpha_{j}=\pm1\}}\left\{ \left(\prod_{j=1}^{n}\alpha_{j}e^{\alpha_{j}i\frac{a\pi\xi}{\beta}}\right)\prod_{j=1}^{n-1}\prod_{k=j+1}^{n}\left(1+i\frac{\alpha_{k}-\alpha_{j}}{2}\frac{\sin\pi\xi}{\sinh\left(\vartheta_{k}-\vartheta_{j}\right)}\right)\right\} \end{eqnarray*}
The parity properties of the form factors are determined by \[
P_{n}^{-a}=\left(-1\right)^{n}P_{n}^{a}\]
As a result we obtain\begin{eqnarray*}
\langle0|\cos\left(a\Phi\right)|B_{1}(\vartheta_{1})\dots B_{1}(\vartheta_{n})\rangle & = & \frac{1+(-1)^{n}}{2}\langle0|\exp\left(ia\Phi\right)|B_{1}(\vartheta_{1})\dots B_{1}(\vartheta_{n})\rangle\\
\langle0|\sin\left(a\Phi\right)|B_{1}(\vartheta_{1})\dots B_{1}(\vartheta_{n})\rangle & = & \frac{1-(-1)^{n}}{2i}\langle0|\exp\left(ia\Phi\right)|B_{1}(\vartheta_{1})\dots B_{1}(\vartheta_{n})\rangle\end{eqnarray*}
Form factors of higher breather can be computed using the bootstrap
fusion rules \cite{smirnov}. To obtain $B_{3}$ one can fuse $2$
$B_{1}$-s into a $B_{2}$ and then a $B_{1}$ and a $B_{2}$ into
a $B_{3}$. The fusion angles and three-particle couplings can be
read off the $B_{1}-B_{1}$ and $B_{1}-B_{2}$ $S$ matrices \cite{zam2}\begin{eqnarray}
S_{11} & = & \frac{\sinh\vartheta+i\sin\pi\xi}{\sinh\vartheta-i\sin\pi\xi}\label{eq:s11_sg}\\
S_{12} & = & \frac{\sinh\vartheta+i\sin\frac{\pi}{2}\xi}{\sinh\vartheta-i\sin\frac{\pi}{2}\xi}\frac{\sinh\vartheta+i\sin\frac{3\pi}{2}\xi}{\sinh\vartheta-i\sin\frac{3\pi}{2}\xi}\nonumber \end{eqnarray}
If the particle $c$ occurs as a bound state of $a$ and $b$ then
the corresponding $S$ matrix has a pole\[
S_{ab}\left(\vartheta\sim iu_{ab}^{c}\right)\sim\frac{i|\gamma_{ab}^{c}|^{2}}{\vartheta-iu_{ab}^{c}}\]
from which we obtain (choosing the three-particle couplings real and
positive)\begin{eqnarray*}
 &  & u_{11}^{2}=\pi\xi\qquad,\qquad\gamma_{11}^{2}=\sqrt{2\tan\pi\xi}\\
 &  & u_{12}^{3}=\frac{3\pi}{2}\xi\qquad,\qquad\gamma_{12}^{3}=\sqrt{2\frac{\tan\pi\xi\tan\frac{3\pi}{2}\xi}{\tan\frac{\pi}{2\xi}}}\end{eqnarray*}
 Therefore the $311$ form factor can be obtained starting from a
$5$-particle $11111$ form factor and fusing the first three $B_{1}$
particles into a $B_{3}$. This means that the decay process $B_{3}\rightarrow B_{1}+B_{1}$
has a non-vanishing amplitude to leading order only if $\delta\neq0$.
Indeed, for $\delta=0$ this process (and its generalization $B_{2n+1}\rightarrow B_{1}+B_{1}$)
is entirely forbidden by the $C$-parity of the breathers. 

Using the notation $x_{j}=\mathrm{e}^{\vartheta_{j}}$ we get the
following result for the $5$-particle form factor\begin{eqnarray}
F_{11111}^{a}\left(\vartheta_{1},\dots,\vartheta_{5}\right) & = & \mathcal{G}_{a}\bar{\lambda}^{5}Q_{11111}^{a}\left(\vartheta_{1},\dots,\vartheta_{5}\right)\prod_{j=1}^{4}\prod_{k=j+1}^{5}R\left(\vartheta_{k}-\vartheta_{j}\right)\label{eq:F11111}\\
Q_{11111}^{a}\left(\vartheta_{1},\dots,\vartheta_{5}\right) & = & -i[a]\left\{ [a]^{4}-[a]^{2}+\frac{\left([a]^{2}(\sigma_{1}\sigma_{4}-\sigma_{5})+\sigma_{5}\right)\left(\sigma_{2}\sigma_{3}-4\sigma_{5}\cos^{2}\pi\xi\right)}{{\displaystyle \prod_{j=1}^{4}\prod_{k=j+1}^{5}(x_{j}+x_{k})}}\right\} \nonumber \end{eqnarray}
where the elementary symmetric polynomials $\sigma_{k}$ are defined
by the generating function\[
{\displaystyle \prod_{j=1}^{5}(x_{j}+x)}=\sum_{k=0}^{5}x^{5-k}\sigma_{k}\]
and \[
[a]=\frac{\sin\frac{\pi\xi a}{\beta}}{\sin\pi\xi}\]

In the following we choose $\delta=-\pi/2$ because it will be convenient
later due to another $Z_{2}$ symmetry which exists at this point
(and which, on the other hand, forbids every process $B_{2n}\rightarrow B_{1}+B_{1}$
to the lowest order). Then the form factors of the perturbing field
$\Psi=\sin\frac{\beta}{2}\Phi$ can be written as

\[
F_{1\dots1}^{\Psi}\left(\vartheta_{1},\dots,\vartheta_{n}\right)=\langle0|\sin\left(\frac{\beta}{2}\Phi\right)|B_{1}(\vartheta_{1})\dots B_{1}(\vartheta_{n})\rangle=-iF_{1\dots1}^{\beta/2}\left(\vartheta_{1},\dots,\vartheta_{n}\right)\]
for $n$ odd. 

First we fuse two $B_{1}$'s to obtain the form factor $F_{2111}$.
The relevant equation is\[
i\gamma_{11}^{2}F_{2111}^{\Psi}\left(\vartheta_{1},\vartheta_{3},\vartheta_{4},\vartheta_{5}\right)=\mathop{\mathrm{Res}}_{\epsilon=0}F_{11111}^{\Psi}\left(\vartheta_{1}+\frac{1}{2}\left(i\pi\xi-\epsilon\right),\vartheta_{1}-\frac{1}{2}\left(i\pi\xi-\epsilon\right),\vartheta_{3},\vartheta_{4},\vartheta_{5}\right)\]
The only singularity comes from the factor $R\left(\vartheta_{2}-\vartheta_{1}\right)$
in (\ref{eq:F11111}). The residue can be calculated quite simply
due to (\ref{eq:minff_defining_relations}) with the result\[
\mathop{\mathrm{Res}}_{\vartheta=-i\pi\xi}R\left(\vartheta\right)=\mathop{\mathrm{Res}}_{\vartheta=-i\pi\xi}S_{11}\left(-\vartheta\right)R\left(-\vartheta\right)=i\left(\gamma_{11}^{2}\right)^{2}R\left(i\pi\xi\right)\]
and we get \begin{eqnarray*}
F_{2111}^{\Psi}\left(\vartheta_{1},\vartheta_{3},\vartheta_{4},\vartheta_{5}\right) & = & \mathcal{G}_{\beta/2}\bar{\lambda}^{5}\gamma_{11}^{2}R\left(i\pi\xi\right)Q_{11111}^{\Psi}\left(\vartheta_{1}+\frac{i\pi\xi}{2},\vartheta_{1}-\frac{i\pi\xi}{2},\vartheta_{3},\vartheta_{4},\vartheta_{5}\right)\times\\
 &  & \prod_{j=3}^{4}\prod_{k=j+1}^{5}R\left(\vartheta_{k}-\vartheta_{j}\right)\prod_{l=3}^{5}R\left(\vartheta_{l}-\vartheta_{1}-\frac{i\pi\xi}{2}\right)R\left(\vartheta_{l}-\vartheta_{1}+\frac{i\pi\xi}{2}\right)\\
\mathrm{where} &  & Q_{11111}^{\Psi}=-iQ_{11111}^{\beta/2}\end{eqnarray*}
Now we fuse $B_{1}$ and $B_{2}$ into $B_{3}$\[
i\gamma_{12}^{3}F_{311}^{\Psi}\left(\vartheta_{1},\vartheta_{4},\vartheta_{5}\right)=\mathop{\mathrm{Res}}_{\epsilon=0}F_{211}^{\Psi}\left(\vartheta_{1}+\frac{1}{2}i\pi\xi-\frac{\epsilon}{2},\vartheta_{1}-i\pi\xi+\frac{\epsilon}{2},\vartheta_{4},\vartheta_{5}\right)\]
Using the same tricks as above plus the identity \[
S_{11}\left(2i\pi\xi\right)=\frac{\tan\frac{3\pi\xi}{2}}{\tan\frac{\pi\xi}{2}}=\left(\frac{\gamma_{12}^{3}}{\gamma_{11}^{2}}\right)^{2}\]
we obtain \begin{eqnarray*}
F_{311}^{\Psi}\left(\vartheta_{1},\vartheta_{4},\vartheta_{5}\right) & = & \mathcal{G}_{\beta/2}\bar{\lambda}^{5}\gamma_{11}^{2}\gamma_{12}^{3}R\left(i\pi\xi\right)^{2}R\left(2i\pi\xi\right)\times\\
 &  & Q_{11111}^{\Psi}\left(\vartheta_{1},\vartheta_{4},i\pi\xi+\vartheta_{5},\vartheta_{5},-i\pi\xi+\vartheta_{5},\right)\times\\
 &  & R\left(\vartheta_{5}-\vartheta_{4}\right)\prod_{j=4,5}\left[R\left(\vartheta_{j}-\vartheta_{1}+i\pi\xi\right)R\left(\vartheta_{j}-\vartheta_{1}\right)R\left(\vartheta_{j}-\vartheta_{1}-i\pi\xi\right)\right]\end{eqnarray*}
To obtain the decay rate we need \[
f_{311}=\left|F_{311}^{\Psi}\left(i\pi,\vartheta_{c},-\vartheta_{c}\right)\right|\]
where the rapidity $\vartheta_{c}$ of the outgoing $B_{1}$ particles
can be computed from\[
2\cosh\left(\vartheta_{c}\right)=\frac{m_{3}}{m_{1}}=\frac{\sin\frac{3\pi\xi}{2}}{\sin\frac{\pi\xi}{2}}\]
The result can be written as \[
f_{311}=\mathcal{G}_{\beta/2}\bar{\lambda}^{5}\gamma_{11}^{2}\gamma_{12}^{3}\mathcal{Q}_{11111}\left(\xi\right)\mathcal{R}_{311}\left(\xi\right)\]
where \begin{eqnarray*}
\mathcal{Q}_{11111}\left(\xi\right) & = & -Q_{11111}^{\Psi}\left(\vartheta_{c},-\vartheta_{c},i\pi(1+\xi),i\pi,i\pi(1-\xi)\right)\\
 & = & \frac{\left(1+2\cos\pi\xi\right)\left(1+2\cos\pi\xi+2\cos2\pi\xi\right)}{64\cos\pi\xi\cos^{5}\frac{\pi\xi}{2}}\end{eqnarray*}
and \begin{eqnarray*}
\mathcal{R}_{311}\left(\xi\right) & = & \Big|R\left(i\pi\xi\right)^{2}R\left(2i\pi\xi\right)R\left(-2\vartheta_{c}\right)\times\\
 &  & R\left(\vartheta_{c}-i\pi(1-\xi)\right)R\left(\vartheta_{c}-i\pi\right)R\left(\vartheta_{c}-i\pi(1+\xi)\right)\times\\
 &  & R\left(-\vartheta_{c}-i\pi(1-\xi)\right)R\left(-\vartheta_{c}-i\pi\right)R\left(-\vartheta_{c}-i\pi(1+\xi)\right)\Big|\end{eqnarray*}
We can make use of eq. (\ref{eq:R_ipi_identity}) to shift the arguments
of the $R(2\vartheta_{c}$), $R(i\pi\xi)$ and $R(2i\pi\xi)$ factors
into the validity range of the integral representation (\ref{eq:R_intrep}).
The result is\[
\mathcal{R}_{311}\left(\xi\right)=\frac{1}{2}\left|\frac{\sinh2\vartheta_{c}}{\sinh2\vartheta_{c}-i\sin\pi\xi}\,\frac{\cos\pi\xi}{2\cos\pi\xi+1}\,\frac{R\left(\vartheta_{c}-i\pi(1-\xi)\right)^{2}R\left(\vartheta_{c}-i\pi\right)^{2}R\left(\vartheta_{c}-i\pi(1+\xi)\right)^{2}}{R\left(i\pi(\xi-1)\right)^{2}R\left(i\pi(2\xi-1)\right)R\left(-2\vartheta_{c}-i\pi\right)}\right|\]
For the dimensionless matrix element $s_{311}$ we obtain\begin{equation}
s_{311}\left(\xi\right)=\frac{f_{311}}{M^{\frac{\xi}{2+2\xi}}}=\tilde{\mathcal{G}}\left(\xi\right)\bar{\lambda}\left(\xi\right)^{5}2\tan\pi\xi\sqrt{\frac{\tan\frac{3\pi\xi}{2}}{\tan\frac{\pi\xi}{2}}}\mathcal{Q}_{11111}\left(\xi\right)\mathcal{R}_{311}\left(\xi\right)\label{eq:s311_result}\end{equation}
where \[
\tilde{\mathcal{G}}\left(\xi\right)=\left[\frac{\sqrt{\pi}\Gamma\left(\frac{1+\xi}{2}\right)}{2\Gamma\left(\frac{\xi}{2}\right)}\right]^{\frac{\xi}{2+2\xi}}\exp\left\{ \int_{0}^{\infty}\frac{dt}{2t}\left[\frac{\sinh\left(\frac{\xi t}{1+\xi}\right)}{\cosh\left(\frac{t}{1+\xi}\right)\sinh t}-\frac{\xi}{1+\xi}e^{-2t}\right]\right\} \]

\section{\label{sec:dd_appendix} Extracting $\Delta\delta=\min\delta_{1}-\max\delta_{2}$
from the {}``mini-Hamiltonian'' }

We start from the expression (\ref{eq:miniham_levels12}) for the
two levels\[
E_{1,2}=m_{c}+A\lambda-\alpha(L-L_{\mathrm{min}})\pm\sqrt{\alpha^{2}(L-L_{\mathrm{min}})^{2}+B^{2}\lambda^{2}}\]
From this the phase shift functions $\delta_{1,2}$ defined in (\ref{eq:BW_phaseshift12def})
can be expressed as functions of $L$\[
\delta_{1,2}(L)=-L\sqrt{\left(\frac{E_{1,2}(L)}{2}\right)^{2}-m_{1}^{2}}\]
We demonstrate how to find the extremal value of $\delta_{1}(L)$.
We solve the equation\[
\frac{d\delta_{1}(L)}{dL}=0\]
by substituting $L=L_{\mathrm{min}}+\kappa\lambda$ and expanding
in $\lambda$. The lowest non-vanishing order is $\lambda^{0}$ and
we find the following equation for $\kappa$\[
m_{c}^{2}-4m_{1}^{2}-L_{0}m_{c}\left(\alpha-\frac{\alpha^{2}\kappa}{\sqrt{B^{2}+\alpha^{2}\kappa^{2}}}\right)=0\]
Solving this equation and substituting the solution back into $\delta_{1}$,
we find that to linear order in $\lambda$ \[
\min\delta_{1}=-L_{0}\sqrt{\left(\frac{m_{c}}{2}\right)^{2}-m_{1}^{2}}+B\lambda\frac{\sqrt{-(m_{c}^{2}-4m_{1}^{2}-2\alpha L_{0}m_{c})}}{2\alpha}\]
Performing a similar calculation for $\delta_{2}$ we obtain\[
\max\delta_{2}=-L_{0}\sqrt{\left(\frac{m_{c}}{2}\right)^{2}-m_{1}^{2}}-B\lambda\frac{\sqrt{-(m_{c}^{2}-4m_{1}^{2}-2\alpha L_{0}m_{c})}}{2\alpha}\]
The difference is \begin{equation}
\min\delta_{1}-\max\delta_{2}=B\lambda\frac{\sqrt{-(m_{c}^{2}-4m_{1}^{2}-2\alpha L_{0}m_{c})}}{\alpha}\label{eq:minihamphasediff}\end{equation}
We can go further by calculating the slope of the phase shift at $\lambda=0$.
The energy of the two-particle level (the one which is not independent
of $L$) is then\begin{equation}
E=m_{c}-2\alpha(L-L_{0})\label{eq:tplevel_expansion}\end{equation}
Using\[
\delta_{\lambda=0}(E)=-L(E)\sqrt{\left(\frac{E}{2}\right)^{2}-m_{1}^{2}}\]
we get \[
\left.\frac{d\delta_{\lambda=0}}{dE}\right|_{E=m_{c}}=\frac{m_{c}^{2}-4m_{1}^{2}-2\alpha L_{0}m_{c}}{4\alpha\sqrt{m_{c}^{2}-4m_{1}^{2}}}\]
which must be negative for the Breit-Wigner method to apply, so eq.
(\ref{eq:minihamphasediff}) makes sense. To leading order in $\lambda$
$\delta_{\lambda=0}$ can be substituted with the background phase
shift $\delta_{0}$ and $m_{c}$ with the resonance position $E_{0}$
and so we obtain\[
\min\delta_{1}-\max\delta_{2}=2B\lambda\sqrt{-\frac{1}{\alpha}\sqrt{m_{c}^{2}-4m_{1}^{2}}\left.\frac{d\delta_{0}}{dE}\right|_{E=E_{0}}}\]

\section{The {}``mini-Hamiltonian'' coefficient $C$}

\subsection{\label{sub:C_yangbethe} Determining $C$ from finite size corrections}

Let us calculate the shift of a two-particle level to first order
in $\lambda$. For zero total momentum, the relations (\ref{eq:Luscher_quant})
can be reformulated as\[
L\sqrt{\left(\frac{E\left(\lambda\right)}{2}\right)^{2}-m_{1}(\lambda)^{2}}+\delta(E\left(\lambda\right),\lambda)=2n\pi\]
To first order in $\lambda$, we get the following equation for the
energy shift $\delta E$: \[
L\frac{\frac{E\delta E}{4}-m_{1}\delta m_{1}}{\sqrt{\left(\frac{E}{2}\right)^{2}-m_{1}^{2}}}+\frac{d\delta_{0}(E)}{dE}\delta E+\frac{\partial\delta(E,\lambda)}{\partial\lambda}\lambda=0\]
where $E$ is the energy, $m_{1}$ is the particle mass and $\delta_{0}(E)$
is the phase shift at $\lambda=0$, which satisfy\begin{equation}
L\sqrt{\left(\frac{E}{2}\right)^{2}-m_{1}^{2}}+\delta_{0}(E)=2n\pi\label{eq:luscher2}\end{equation}
and $\delta m_{1}$ is the mass shift to first order in $\lambda$.
As a result\[
\delta E=\frac{Lm_{1}\delta m_{1}-\sqrt{\left(\frac{E}{2}\right)^{2}-m_{1}^{2}}\frac{\partial\delta(E,\lambda=0)}{\partial\lambda}\lambda}{\frac{LE}{4}+\sqrt{\left(\frac{E}{2}\right)^{2}-m_{1}^{2}}\frac{d\delta_{0}(E)}{dE}}\]
Using (\ref{smatr_corr}) we can write at the level crossing $L=L_{0}$
(recall that $E(L_{0})=m_{c}$)\begin{eqnarray*}
-\sqrt{\left(\frac{m_{c}}{2}\right)^{2}-m_{1}^{2}}\left.\frac{\partial\delta(m_{c},\lambda)}{\partial\lambda}\right|_{\lambda=0} & = & i\mathrm{e}^{-i\delta_{0}(m_{c})}\sqrt{\left(\frac{m_{c}}{2}\right)^{2}-m_{1}^{2}}\left.\frac{\partial S_{11}(m_{c},\lambda)}{\partial\lambda}\right|_{\lambda=0}\\
 & = & m_{1}\sinh\vartheta_{1}^{(c11)}\frac{S_{11}\left(-2\vartheta_{1}^{(c11)}\right)F_{1111}^{\Psi}\left(i\pi,\,2\vartheta_{1}^{(c11)}+i\pi,\,0,\,2\vartheta_{1}^{(c11)}\right)}{m_{1}^{2}\sinh2\vartheta_{1}^{(c11)}}\\
 & = & \frac{F_{1111}^{\Psi}\left(\vartheta_{1}^{(c11)}+i\pi,-\vartheta_{1}^{(c11)}+i\pi,-\vartheta_{1}^{(c11)},\vartheta_{1}^{(c11)}\right)}{m_{c}}\end{eqnarray*}
From (\ref{mass_correction}) \[
m_{1}\delta m_{1}=\lambda F_{11}^{\Psi}(i\pi,0)\]
On the other hand, using $E(L)=2\sqrt{p(L)+m_{1}^{2}}$ (with the
two particles having momentum $\pm p(L)$) we get from (\ref{eq:luscher2})\[
\frac{LE}{4}+\sqrt{\left(\frac{E}{2}\right)^{2}-m_{1}^{2}}\frac{d\delta_{0}(E)}{dE}=-\frac{pE}{4}\frac{dL}{dp}\quad\mathop{\rightarrow}_{L=L_{0}}\quad\frac{m_{c}}{4\left.\left(-\frac{1}{p}\frac{dp(L)}{dL}\right)\right|_{L=L_{0}}}\]
and so we obtain\[
C=\left.\left(-\frac{1}{p}\frac{dp(L)}{dL}\right)\right|_{L=L_{0}}\left(\frac{4F_{1111}^{\Psi}\left(\vartheta_{1}^{(c11)}+i\pi,-\vartheta_{1}^{(c11)}+i\pi,-\vartheta_{1}^{(c11)},\vartheta_{1}^{(c11)}\right)}{m_{c}^{2}}+L_{0}\frac{4F_{11}^{\Psi}(i\pi,0)}{m_{c}}\right)\]

\subsection{\label{sub:C_disconnected} Disconnected parts of the four-particle
matrix element}

We need to determine the matrix element $\langle A_{1}(p)A_{1}(-p)|\Psi(0)|A_{1}(p)A_{1}(-p)\rangle_{L=L_{0}}$.
Returning to rapidity variables and using the crossing properties
of form factors we can write (in infinite volume)\begin{eqnarray*}
\,_{\mathrm{in}}\langle A_{1}(\vartheta_{3})A_{1}(\vartheta_{4})|\Psi(0)|A_{1}(\vartheta_{1})A_{1}(\vartheta_{2})\rangle_{\mathrm{in}} & = & F_{1111}^{\Psi}\left(\vartheta_{3}+i\pi,\vartheta_{4}+i\pi,\vartheta_{1},\vartheta_{2}\right)+\\
 &  & 2\pi\delta(\vartheta_{1}-\vartheta_{3})F_{11}\left(\vartheta_{4}+i\pi,\vartheta_{2}\right)+\\
 &  & 2\pi\delta(\vartheta_{2}-\vartheta_{4})F_{11}\left(\vartheta_{3}+i\pi,\vartheta_{1}\right)+\\
 &  & 2\pi\delta(\vartheta_{1}-\vartheta_{3})\delta(\vartheta_{2}-\vartheta_{4})\langle0|\Psi(0)|0\rangle\end{eqnarray*}
The last term can be dropped because it is related to the vacuum energy
shift, but we normalized the finite volume energy levels by subtracting
the ground state. In \cite{nonintegrable}, the other two disconnected
pieces were canceled by mass shift counter terms; however, the finite
volume Hamiltonian does not contain such terms. The $\delta$ functions
can be written as \[
2\pi\delta\left(\vartheta_{1}-\vartheta_{2}\right)=2\pi\delta(p_{1}-p_{2})m_{1}\cosh\vartheta_{1}\]
and in finite volume ($L_{0}$)\[
\left.2\pi\delta\left(\vartheta_{1}-\vartheta_{2}\right)\right|_{\vartheta_{1}=\vartheta_{2}}=L_{0}m_{1}\cosh\vartheta_{1}\]
We need the matrix element at the special rapidity values $\vartheta_{1}=-\vartheta_{2}=\vartheta_{3}=-\vartheta_{4}=\vartheta_{1}^{(c11)}$
(recall that $m_{c}=2m_{1}\cosh\vartheta_{1}^{(c11)}$) which gives\begin{eqnarray*}
\langle A_{1}(p)A_{1}(-p)|\Psi(0)|A_{1}(p)A_{1}(-p)\rangle_{L=L_{0}} & = & F_{1111}^{\Psi}\left(\vartheta_{1}^{(c11)}+i\pi,-\vartheta_{1}^{(c11)}+i\pi,-\vartheta_{1}^{(c11)},\vartheta_{1}^{(c11)}\right)+\\
 &  & m_{c}L_{0}F_{11}^{\Psi}\left(i\pi,0\right)\end{eqnarray*}

\end{document}